# Measurements of extremely low radioactivity levels in BOREXINO

BOREXINO Collaboration

C. Arpesella[a], H.O. Back[b], M. Balata[a], T. Beau[o], G. Bellini[l&], J. Benziger[r], S. Bonetti[l], A. Brigatti[l], C. Buck[h], B. Caccianiga[l], L. Cadonati[s], F. Calaprice[s], G. Cecchet[p], M. Chen[i], O. Dadoun[o], D. D'Angelo[g], A. De Bari[p], A. de Bellefon[o], E. De Haas[s], H. de Kerret[o], A. Derbin[e§], M. Deutsch[d], A. Di Credico[a], F. Elisei[q], A. Etenko[m], F. von Feilitzsch[f], R. Fernholz[s], R. Ford[as], D. Franco[l], B. Freudiger[h], C. Galbiati[s$], F. Gatti[g], S. Gazzana[a], M.G. Giammarchi[l], D. Giugni[l], M. Göger-Neff[f], T. Goldbrunner[f], A. Golubchikov[l], A. Goretti[l], C. Grieb[f], C. Hagner[b], T. Hagner[f], W. Hampel[h*], E. Harding[s], F.X. Hartmann[bh], R. von Hentig[f], G. Heusser[h], M. Hult[j], A. Ianni[ls], A.M. Ianni[s], J. Kiko[h], T. Kirsten[h], M. Köhler[j], G. Korga[l¶], G. Korschinek[f], Y. Kozlov[m], D. Kryn[o], P. LaMarche[ls], M. Laubenstein[a], C. Lendvai[f], F. Loeser[s], P. Lombardi[l], K. McCarty[s], I. Machulin[m], S. Malvezzi[l], J. Maneira[l], I. Manno[c], G. Manuzio[g], A. Martemianov[a†], F. Masetti[q], U. Mazzucato[q], E. Meroni[l], L. Miramonti[l], M.E. Monzani[a], P. Musico[g], H. Neder[h], L. Niedermeier[f], S. Nisi[a], L. Oberauer[a‡#], M. Obolensky[o], F. Ortica[q], M. Pallavicini[g], L. Papp[l¶], L. Perasso[l], A. Pocar[s], R.S. Raghavan[n], G. Ranucci[l], W. Rau[ah], A. Razeto[g], E. Resconi[g], T. Riedel[f], A. Sabelnikov[l†], C. Salvo[g], R. Scardaoni[l], S. Schönert[h], K.H. Schuhbeck[f], T. Shutt[s], H. Simgen[h], A. Sonnenschein[s], O. Smirnov[e], A. Sotnikov[e], M. Skorokhvatov[m], S. Sukhotin[m], V. Tarasenkov[m], R. Tartaglia[a], G. Testera[g], P.R. Trincherini[t], V. Vyrodov[m], R.B. Vogelaar[b], D. Vignaud[o], S. Vitale[g], M. Wójcik[k], O. Zaimidoroga[e], G. Zuzel[k].

*a) Laboratori Nazionali del Gran Sasso - Assergi - Italy*
*b) Virginia Polytechnic Institute and State University - Blacksburg VA - USA*
*c) Research Institute for Particle and Nuclear Physics- Budapest - Hungary*
*d) Massachusetts Institute of Technology - Cambridge MA - USA*
*e) Joint Institute for Nuclear Research - Dubna - Russia*
*f) Technical University Munich - Garching - Germany*
*g) Physics Department of the University and INFN - Genova - Italy*
*h) Max-Planck-Institut für Kernphysik - Heidelberg - Germany*
*i) Queen's University - Kingston - Canada*
*j) Institute for Reference Materials and Measurements – Geel - Belgium*
*k) Jagellonian University - Cracow - Poland*
*l) Physics Department of the University and INFN - Milano - Italy*
*m) Kurchatov Institut - Moscow - Russia*
*n) Bell Laboratories - Murray Hill NJ - USA*
*o) Collège de France - Paris - France*
*p) Physics Department of the University and INFN - Pavia - Italy*
*q) Chemistry Department of the University and INFN - Perugia - Italy*
*r) Department of Chemical Engineering, Princeton University - Princeton NJ - USA*
*s) Department of Physics, Princeton University - Princeton NJ - USA*
*t) Joint Research Center of EU - Environment Institute - Ispra - Italy*

*&) Spokesperson*
*#) Project manager*
*$) INFN fellowship*
*§) on leave of absence from Gatchina Institute, S. Petersburg, Russia*
*¶) on leave of absence from c)*
*†) on leave of absence from m)*
*‡) on leave of absence from f)*

*\*) Corresponding author (E-mail address:Wolfgang.Hampel@mpi-hd.mpg.de)*

**Abstract**

The techniques researched, developed and applied towards the measurement of radioisotope concentrations at ultra-low levels in the real-time solar neutrino experiment BOREXINO at Gran Sasso are presented and illustrated with specific results of widespread interest. We report the use of low-level germanium gamma spectrometry, low-level miniaturized gas proportional counters and low background scintillation detectors developed in solar neutrino research. Each now sets records in its field. We additionally describe our techniques of radiochemical ultra-pure, few atom manipulations and extractions. Forefront measurements also result from the powerful combination of neutron activation and low-level counting. Finally, with our techniques and commercially available mass spectrometry and atomic absorption spectroscopy, new low-level detection limits for isotopes of interest are obtained.

# 1 Introduction

The BOREXINO experiment [1] aims to measure low energy solar neutrinos in real time by elastic neutrino-electron scattering. The mono-energetic neutrinos from $^7$Be at 862 keV are of most interest. The experiment, located near L'Aquila in Italy in the underground laboratory LNGS (Laboratori Nazionali del Gran Sasso) will use 300 t of liquid scintillator (fiducial volume 100 t) to detect the scattered electrons. The energy window in which the $^7$Be neutrino induced events will be observed is in the range of 250 to 800 keV, at an expected rate of a few tens of events per day. Since these very rare events are practically indistinguishable from other ionizing events produced by natural radioactivity at the same energy, extremely high radiopurity standards must be met by the experiment.

Muon induced background events are strongly suppressed by the overburden (~3500 meter water equivalent, m.w.e.) and by the aid of an additional muon veto shield. The shell type structure of the detector, as shown in Figure 1, will reduce the environmental γ-ray and neutron fluxes to an insignificant level at the central fiducial volume. The total shield thickness of ~5 m.w.e. consists of ~2.1 m pure water, ~2.6 m Pseudocumene (1,2,4-trimethylbenzene, PC, $\rho \cong 0.89$ g/cm$^3$) and the outer ~1.2 m of the liquid scintillator. The latter is contained in a thin nylon wall inner vessel. The scintillation light is viewed by 2200 photo multiplier tubes (PMT) mounted on the stainless steel sphere (SSS). A muon veto is formed by 200 outward looking PMTs detecting the Čerenkov light of through-going muons in the water buffer. The outer shell of the detector is a stainless steel tank of 18 m diameter. The inner part of the PC buffer is protected against radon ($^{222}$Rn) emanated from the PMTs and the SSS by another nylon film (outer vessel). An elaborate system handles and purifies some 300 tons of PC based scintillator and ~1040 tons of PC buffer liquid. Further ancillary plants are a water purification system and a nitrogen plant to supply the experiment with high purity water and nitrogen. BOREXINO is now in the final phase of construction in hall C of the LNGS.

In this paper we describe the analytical procedures and methods applied by the BOREXINO collaboration in order to control and monitor the required radiopurity in the construction of the detector itself, its peripheral sub-systems and their operation. After a discussion of the radiopurity requirements for BOREXINO (section 2), results obtained with the following techniques are presented: low-level germanium γ–ray spectrometry (section 3), low-level radon counting (section 4), neutron activation analysis (section 5), inductively coupled plasma mass spectrometry (section 6) and graphite furnace atomic absorption spectroscopy (section 7). Finally, some results on extreme low-level measurements performed with the Counting Test Facility (CTF, the BOREXINO prototype experiment) are included here for completeness (section 8), although already published elsewhere [2-4].

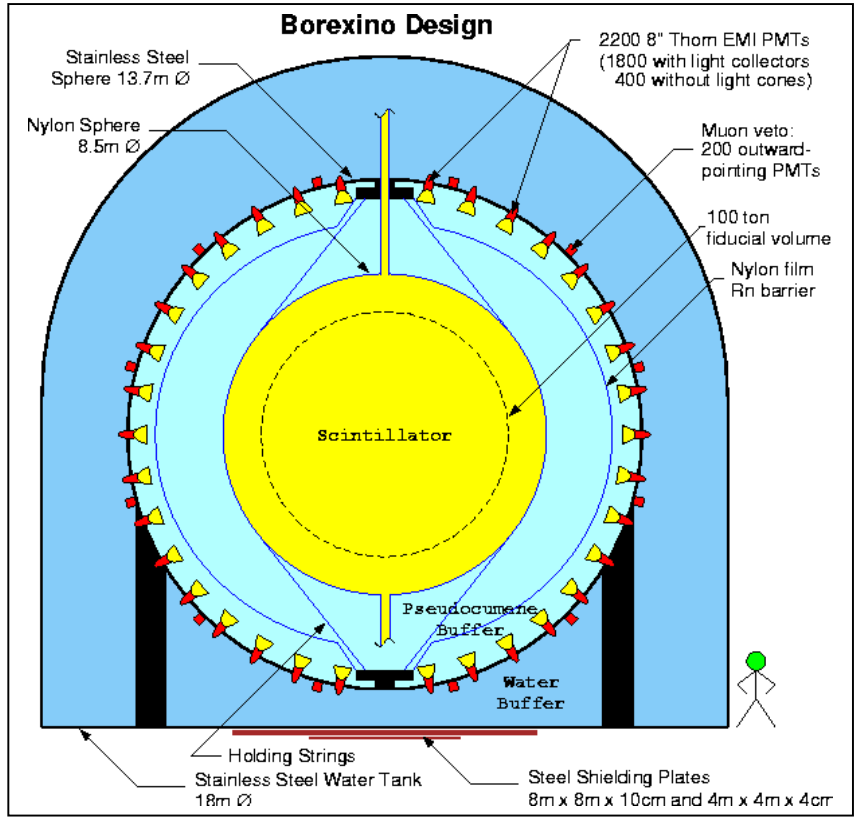

Figure 1: Design of the BOREXINO Solar Neutrino Detector. The 300 tons of scintillator contained in the inner nylon vessel are shielded by 1040 tons of buffer liquid (pseudocumene) which fills the outer stainless steel sphere (which also supports the array of the 2200 photomultiplier tubes). The whole arrangement is immersed in high purity water contained in the main tank.

## 2   Radiopurity requirements

The most important sources of contamination for BOREXINO can be classified according to their origin:

    Primordial radionuclides, such as $^{238}$U, $^{232}$Th (and their decay chains) and $^{40}$K.
    Cosmogenic radionuclides, such as $^{7}$Be and $^{14}$C.
    Anthropogenic (man-made) radionuclides, such as $^{60}$Co, $^{85}$Kr and $^{137}$Cs.

Due to their environmental occurrence and activity, the isotopes $^{40}$K, $^{232}$Th, and $^{238}$U are the most prominent contaminants. Table 1 lists the design goals for the radiopurity concentrations allowed for the major components of the detector. As can be seen, the tolerable activity concentrations strongly decrease from the outer to the inner parts of the detector. This trend is also conform with the variation of the mass fractions of the components. The main concerns in the fiducial volume are α, β and γ activities, whereas from the periphery only the γ activity poses a potential hazard. For simplicity in the notation, the U/Th concentrations given in Table 1 assume secular equilibrium only. For cases where the $^{222}$Rn diffusion is important, it is stated separately. The purity regimes for the scintillator are far beyond the sensitivity of any low-level counting technique available so far and therefore represented the major challenge of the experiment.

In some parts of this report the concentrations of primordial nuclides are given as mass fractions (g/g), like in this table, but in others activity concentrations (e.g. Bq/kg) are used. Therefore we give here the conversion factors for the case of secular equilibrium:

  U: $10^{-9}$ g/g = 12.35 mBq/kg;  Th: $10^{-9}$ g/g = 4.06 mBq/kg;  $^{nat}$K: $10^{-9}$ g/g = 31 µBq/kg.

Table 1: The design radiopurities of the major BOREXINO detector components.

| Components | $^{238}$U (g/g) | $^{232}$Th (g/g) | $^{nat}$K (g/g) | other |
|---|---|---|---|---|
| Stainless steel tank wall | $1\cdot10^{-8}$ | $1\cdot10^{-8}$ | $1\cdot10^{-5}$ | |
| Water buffer | $1\cdot10^{-10}$ | $1\cdot10^{-10}$ | $1\cdot10^{-7}$ | |
| Water CTF shield | $1\cdot10^{-13}$ | $2\cdot10^{-13}$ | $2\cdot10^{-10}$ | $^{222}$Rn – 1 μBq/kg |
| Stainless Steel Sphere | $2\cdot10^{-10}$ | $1\cdot10^{-10}$ | $3\cdot10^{-7}$ | |
| PMTs | $3\cdot10^{-8}$ | $1\cdot10^{-8}$ | $2\cdot10^{-5}$ | |
| PC buffer | $1\cdot10^{-15}$ | $1\cdot10^{-15}$ | $5\cdot10^{-12}$ | |
| Nylon film for inner vessel | $5\cdot10^{-12}$ | $2\cdot10^{-11}$ | $1\cdot10^{-8}$ | $^{226}$Ra – 12 μBq/kg |
| Scintillator | $1\cdot10^{-16}$ | $1\cdot10^{-16}$ | $1\cdot10^{-14}$ | $^{14}$C/$^{12}$C – $1\cdot10^{-18}$ |
| N$_2$ for scintillator sparging | | | | $^{222}$Rn – 1 μBq/m$^3$ |

Most modern methods of material processing will disturb the secular equilibrium in the decay chains if the chemistry of the atoms is different as far as the processing is concerned. Thus, for the cases of the natural decay chains, information on deviations from the secular equilibrium is important. This is especially true for the cases in which some sub-series γ-emitting nuclides dominate while the γ-intensity for the first member of the series is low. On the other hand, the first member of the series may be of particular worry even if the γ intensity or energy is low as one nears the center (scintillator) of the detector. Finally, as we will see below, there exists no analytical method which provides by itself complete information for the background contribution from the entire decay chain, so that in some cases different methods for determining it have to be applied.

Special decay features in a decay series, such as the $^{214}$Bi-$^{214}$Po decay pair in the $^{238}$U decay chain, offer a tagging possibility through the exploitation of the short half-life of the daughter nuclide ($T_{1/2}$ for $^{214}$Po is 164 μsec in this case). Such delayed coincidence tags are used in the CTF measurements (see section 8) and they are incorporated into the background software cuts for the full scale experiment. Another special tool of this kind is the discrimination between α and β particles afforded through the use of the scintillation light pulse shape.

The omnipresent $^{222}$Rn, with its diffusive ability as a rare gas, is of special relevance to BOREXINO for several reasons: (i) it is highly soluble in the liquid scintillator; (ii) the short-lived daughter nuclides are energetic γ-ray emitters; (iii) the longer-lived decay product $^{210}$Pb contributes to the background in the $^7$Be neutrino window (because its first daughter nuclide $^{210}$Bi decays by emission of β-particles with energies up to 1.17 MeV endpoint energy and because of the quenched energy signal from the α decay of its second daughter nuclide $^{210}$Po); and finally, (iv) it may emanate from its likewise omnipresent progenitor, $^{226}$Ra. The latter requires us to control the $^{226}$Ra contamination also, especially in regions which are in contact or close to the scintillator. Because of many chemical differences between $^{226}$Ra and some of its progenitors, it is not expected to be in equilibrium with them. In addition, since $^{210}$Pb can accumulate on the materials, limits must be set on the exposure of these regions to ambient $^{222}$Rn in the air.

The $^{14}$C abundance in the organic scintillator determines the energy threshold of the $^7$Be neutrino window. If the latter is set to an energy of 250 keV, as currently planned, then the requirement for the upper limit for the $^{14}$C content of the scintillator is lower than the levels ever achieved. This requirement is set because pile-up along with energy resolution limitations can lead to a "tail" from the $^{14}$C energy spectrum that extends well beyond the maximum β energy of 156 keV and into the neutrino event window at higher energy.

To measure such a low concentration of $^{14}$C and to also study the abundance of U and Th in a mid-scale (5 t) liquid scintillator detector, a prototype for the BOREXINO experiment called the Counting Test Facility (CTF) was constructed at LNGS [2-4]. The CTF is a simplified version of BOREXINO, with three components: an inner vessel containing the scintillator, an open structure holding 100 photomultiplier tubes (PMT) and an outer water buffer tank (11m diameter, 10m height).

Recently, a $^{222}$Rn barrier has been added in front of the PMTs. The shielding against the outside γ-ray flux by ~4.5 m water is comparable to that of the BOREXINO detector (~5 m), however the other shielding volumes are absent. Since in the CTF the scintillator is in direct contact with the shielding water, separated only by a 500 μm thick nylon balloon, the purity requirements of the water in the CTF are about a factor of 100 more severe than for the main BOREXINO detector (see Table 1).

## 3   Low-level germanium γ-ray spectrometry

Almost all solid detector materials, beginning with the concrete components for the base of the 18.5 m diameter tank to the nylon ingots used in the fabrication of the parts of the inner vessel, have been or are planned to be screened by γ-ray spectrometry with germanium detectors. This technique is also widely applied in the selection of materials which are needed for the other analytical procedures.

Low-level Ge spectrometry, if operated underground, can reach sensitivities to levels in the mBq range. For kilogram sized samples the technique reaches sensitivities of mBq/kg which corresponds to ~$10^{-10}$ g/g levels of U or Th contamination. This level of sensitivity is sufficient to control materials or components close to the inner detector (see Table 1). Since the tolerable contamination level at these locations is defined by the γ activity, Ge detectors are ideal for this purpose. A further advantage is, that most measurements can be performed in a nondestructive way without laborious sample pre-treatment.

The most abundant contaminants, $^{40}$K and a large part of the U/Th-decay families, are detectable by their emitted γ-rays. Within certain limits, deviations from secular equilibrium in the U/Th series can also be observed. Most important, this is possible in the case of the $^{232}$Th/$^{228}$Ra/$^{228}$Ac/$^{228}$Th sub-series, where the time for noticeable changes in the activity ratios (and re-establishment of equilibrium) is well within the range of the lifetime of the BOREXINO experiment. This fact is important in light of one major aim of BOREXINO, which is to look for time variations of the solar neutrino flux, especially on a seasonal basis.

For the determinations of the U and Th chain activities, the highest sensitivity is obtained using the sub-chains starting with $^{226}$Ra (U) and $^{228}$Th (Th) due to the high γ-abundance in the decay of their daughters. For the case of solid samples, which are not very fine grained (in which case $^{222}$Rn cannot diffuse out), equilibrium with the rest of the chain can be assumed with the exception of $^{210}$Pb.

### 3.1   The germanium γ-ray spectrometers

Low-level Ge spectrometry for BOREXINO has been performed at three different underground laboratories: LNGS, HADES (High Activity Disposal Experimental Site located in Belgium) and the Max-Planck-Institut für Kernphysik (MPI) in Heidelberg. The attenuation of the muon flux by the rock shielding (m.w.e.) compared to sea level varies from $10^6$ (3500 m.w.e.) at LNGS, $10^4$ (500 m.w.e.) at HADES and 3 (15 m.w.e.) at MPI. The smaller overburden at MPI is partly compensated by cosmic ray veto shields, with which each spectrometer is equipped. A brief description of the individual sites follows.

**LNGS**: The low background germanium counting facility is located underground in a prefabricated building along the by-pass in front of Hall C (across from the location of the BOREXINO detector). The radioactivity in the laboratory's concrete covered rock walls is rather low [5] and the neutron flux underground is three orders of magnitude lower than aboveground. Continuous ventilation of the building with air from outside of the tunnel maintains a $^{222}$Rn concentration at about 30 Bq/m$^3$. The facility is equipped with large volume HP coaxial Ge detectors having high resolution and low intrinsic background. One of them (called GePv) has been installed by the BOREXINO collaboration [6], another one (called GeMPI) by the MPI [7], the other detectors (GsOr, GePaolo and GeCris), partly also used for measurements of BOREXINO samples, belong to other experiments at LNGS.

The germanium detectors have been constructed by Ortec (U.S.A.) and Canberra (Belgium); all the materials used for the cryostat and its internal fittings have been carefully selected. Furthermore the company assembly and mounting procedures of the detectors have been supervised by members of the collaboration and of the Gran Sasso Laboratory staff. GeMPI was assembled at Canberra (Belgium) solely from components fabricated at the MPI with the exception of the crystal itself and the preamplifier-chain. The characteristics of the detectors used in the BOREXINO project are reported in Table 2, together with those of the other facilities.

Each detector is shielded by 25 cm of lead that has a $^{210}$Pb specific activity of 20 Bq/kg. The X-rays and the bremsstrahlung from the lead are stopped by a 10 cm layer of high-conductivity, oxygen-free copper, installed close to the detector. At the bottom, a 5 cm acrylic layer and a cadmium foil reduces the neutron flux. GeMPI is shielded with 20 cm of lead with decreasing $^{210}$Pb concentration (5 Bq/kg for the innermost 5 cm layer) and 5 cm specially selected high purity copper. For $^{222}$Rn protection all shields are housed in airtight boxes of acrylic or steel and are continuously flushed with nitrogen gas. An airlock system allows one to load samples into the shield of GeMPI without $^{222}$Rn exposure of the interior. The sample chamber of GeMPI is especially large so as to accommodate larger samples up to a size of 15 l in volume.

**HADES:** The Institute for Reference Materials and Measurements (IRMM) operates several Ge-spectrometers [8] at the 225 m deep HADES Underground Research Facility (URF), an installation of the SCK/CEN Nuclear Research Center at Mol, Belgium. Access to the facility is through a shaft, which was drilled through the overburden sand and clay. HADES is an experimental facility to study disposal of nuclear waste. The walls are made of thick cast-iron although not particularly selected for low-activity purpose. The IRMM detectors are placed in a part of the shaft which is far away from any hot source experiments. A high ventilation rate keeps the $^{222}$Rn level to an average of 7 Bq/m$^3$. The two detectors applied to BOREXINO measurements were manufactured according to low-level specifications by PGT (Germany) for "Ge-1", and by Eurisys (France) for "Ge-3". They are shielded by 6 cm and 14 cm of electrolytic copper respectively and with 14 to 15 cm of low activity lead (20 Bq/kg and < 2 Bq/kg respectively). The open space around the sample is filled with copper and Teflon to expel $^{222}$Rn.

**MPI:** The low-level laboratory of the MPI für Kernphysik in Heidelberg is easily accessible from the basement of the Gentner Laboratory. The dirt and rock overburden of 15 m.w.e. hardly reduces the muon flux but it does serve to strongly suppress the cosmic-ray secondary neutrons. Active veto detectors around the detectors further reduce the muon induced background of the shielded Ge spectrometers by one or two orders of magnitude [9]. The detector cryostat systems were constructed from pre-selected materials in close collaboration with Canberra (Belgium) for the case of "Ge-II" and DSG (Germany) for the case of "Ge-IV". External γ-rays are shielded by 15 cm of lead having decreasing $^{210}$Pb content (down to 1 Bq/kg) and variable inner linings, made mostly out of thin copper. $^{222}$Rn is expelled from the shield by N$_2$-flushing, by evacuating the sample chamber (for the case of Ge-IV only) and by subsequent filling of the chamber with nitrogen. Ventilation of the room with outside air keeps the $^{222}$Rn at an average level of 30 Bq/m$^3$. A motor generator supplies clean electric power to the low-level laboratory.

In Table 2 the characteristics of all the germanium spectrometers are summarized, including the background rates in the energy windows corresponding to the main peaks of interest: 352 keV – $^{214}$Pb (U); 583 keV – $^{208}$Tl (Th) and 1461 keV – $^{40}$K.

The total background rate shown in the table is a measure of the continuous background below the peaks. It has to be subtracted by interpolation from both sides of the peaks to obtain the net peak count rate. The total background rates as well as the peak background rates of all detectors are much lower than is normally the case for the usual spectrometers (such as those used, for instance, in radiation protection measurements).

Table 2: Counting performance of the germanium spectrometers. The background rates are given in counts per day. Upper limits correspond to a 90% confidence level.

| Detector | Volume (cm$^3$) | Relative efficiency (%)$^a$ | Total and Peak background count rate [d$^{-1}$] | | | |
|---|---|---|---|---|---|---|
| | | | 60-2700 keV | 352 keV | 583 keV | 1461 keV |
| GePv | 363 | 91 | 951 | 5.6 | 3.7 | 6.4 |
| GePaolo | 518 | 113 | 624 | 2.6 | 0.9 | 4.7 |
| GsOr | 414 | 96 | 980 | 3.0 | 2.1 | 8.7 |
| GeCris | 468 | 120 | 221 | < 0.9 | < 0.6 | 2.9 |
| GeMPI | 413 | 102 | 70 | < 0.3 | < 0.2 | 0.7 |
| Ge-1 | 91 | 20 | 404 | 0.8 | 0.6 | 1.9 |
| Ge-3 | 234 | 60 | 359 | 1.1 | < 0.5 | 2.0 |
| Ge-II | 120 | 22 | 3200 | < 1.7 | 2.2 | 6.5 |
| Ge-IV | 170 | 37 | 1400 | 5.8 | 6.0 | 2.5 |

$^a$ relative to a 3" x 3" NaI crystal.

The GeMPI detector is currently the lowest level Ge detector running world-wide. It represents the progress in technology that is also being made in addressing the low-levels needed in the solar neutrino experiments. Consequently GeMPI is used for very critical measurements such as those where the primordial radionuclides have to be checked. It is especially sensitive if enough material exists so that the whole sample chamber can be filled.

## 3.2 Data analysis and results

Each laboratory uses its own spectrum analyzing program, tuned to low count rates. The counting efficiency is determined from either calibrated geometries or by using Monte Carlo simulations. At LNGS and HADES the Monte Carlo code EGS4 (Electrons Gamma Shower) is used. This code has been tested with γ sources and has been found to be accurate to the level of 10-15% (LNGS), depending on the geometry, and to the level of 4 - 10% (HADES). MPI uses the CERN codes Geant 3.21 for GeMPI as well as for the detectors at Heidelberg. For volumetric samples, the simulations give results that agree to better than 5 % for GeMPI, and to better than 10% for the others (with the use of their respective radioisotope spiked mockups). The latest version of Geant (Geant 4) is now implemented at LNGS and at Heidelberg.

Prior to the measurement, the samples are cleaned, if possible, using a solution containing a few percent of nitric acid in water and a solution containing ethyl alcohol. Nonetheless, the first few hours of a measurement are still not taken into account because the results can be influenced by $^{222}$Rn daughter nuclides that have plated out onto the components.

A compilation of selected results obtained with these Ge detectors during the CTF and BOREXINO construction phase is illustrated in Table 3. Only data concerning materials of a more general importance or wider spread interest are reported here. By far, most of the measurements were performed at LNGS. Thus fewer measurements as performed at the other two sites are included.

The $^{226}$Ra activity concentration is determined by the γ-ray lines of $^{214}$Pb and $^{214}$Bi, that of $^{228}$Th by the γ-ray lines of $^{212}$Pb, $^{208}$Tl and partly of $^{228}$Ac. For easier comparison with other measurements in this paper the activities are also given as U, Th and $^{nat}$K concentrations. For the conversion of $^{226}$Ra into U and of $^{228}$Th into Th, secular equilibrium had to be assumed, although there are indications in some cases that this is not justified. Generally the conversion from the original measured activity concentration has to be done with precaution, since equilibrium breaking (e.g. between $^{238}$U and $^{226}$Ra or $^{228}$Ra and $^{228}$Th) is more often the rule for the investigated materials than the exception. The quoted errors represent 1σ statistical uncertainties, the upper limits are given at a 95 % confidence level.

Table 3: Compilation of selected results obtained with Ge detectors (see text for details).

| | $^{226}$Ra (mBq/kg) | $^{238}$U (g/g) | $^{228}$Th (mBq/kg) | $^{232}$Th (g/g) | $^{40}$K (mBq/kg) | $^{nat}$K (g/g) | $^{60}$Co (mBq/kg) |
|---|---|---|---|---|---|---|---|
| **PMT inner parts** | | | | | | | |
| Ceramic plates for dynodes structure | 170± 50 | $(1.4±0.4)\ 10^{-8}$ | 310±60 | $(8±1)\ 10^{-8}$ | 960±450 | $(3±1)\ 10^{-5}$ | < 40 |
| Dynodes | < 280 | $< 2.3\ 10^{-8}$ | 450±163 | $(1.1±0.4)\ 10^{-7}$ | < 240 | $< 7.6\ 10^{-6}$ | < 290 |
| Aluminum for dynodes structure | 1190±100 | $(9.6±0.8)\ 10^{-8}$ | 980±80 | $(2.4±0.2)\ 10^{-7}$ | 2800±600 | $(9±2)\ 10^{-5}$ | < 80 |
| Metallic comp. for dynodes structure | < 62 | $< 5\ 10^{-9}$ | < 20 | $< 5\ 10^{-9}$ | < 340 | $< 1.1\ 10^{-5}$ | < 12 |
| **Glass** | | | | | | | |
| Sand for glass manufacturing | 40±3 | $(3.2±0.3)\ 10^{-9}$ | <3.1 | $< 7.6\ 10^{-10}$ | < 25 | $< 8.1\ 10^{-7}$ | < 1.6 |
| ETL low radiation glass | 820±230 | $(6.6±1.9)\ 10^{-8}$ | 130±12 | $(3.2±0.3)\ 10^{-8}$ | 500 ±120 | $(1.6±0.4)\ 10^{-5}$ | |
| Base glass | 520±90 | $(4.2±0.7)\ 10^{-8}$ | 410±90 | $(1±0.2)\ 10^{-7}$ | 226000±6200 | $(7.3±0.2)\ 10^{-3}$ | |
| **PMT ancillary parts** | | | | | | | |
| Mu-metal | 57±20 | $(5±2)\ 10^{-9}$ | < 27 | $< 6.6\ 10^{-9}$ | < 180 | $< 5.8\ 10^{-6}$ | < 9 |
| Phenolic resin coated sample | 86±12 | $(7±1)\ 10^{-9}$ | 610±203 | $(1.5±0.5)\ 10^{-8}$ | < 400 | $< 1.3\ 10^{-5}$ | < 10 |
| Voltage divider printed circuit board | 170±60 | $(1.4±0.5)\ 10^{-8}$ | 80±40 | $(1.9±1.0)\ 10^{-8}$ | 770±360 | $(2.5±1.2)\ 10^{-5}$ | |
| Complete voltage divider | 680±30 | $(5.5±0.3)\ 10^{-8}$ | 320±20 | $(7.9±0.6)\ 10^{-8}$ | 3200±320 | $(1.0±0.1)\ 10^{-4}$ | < 15 |
| Master Bond EP45HT for PMT sealing | < 40 | $< 3\ 10^{-9}$ | < 24 | $< 6\ 10^{-9}$ | < 310 | $< 1\ 10^{-5}$ | < 11 |
| **Cable and connectors** | | | | | | | |
| Complete RG213 Suhner cable | 22±6 | $(1.8±0.5)\ 10^{-9}$ | < 20 | $< 5\ 10^{-9}$ | < 140 | $< 5\ 10^{-6}$ | |
| Complete Jupiter connector | 56±15 | $(4.5±1.2)\ 10^{-9}$ | 2.4±1.2 | $(6±3)\ 10^{-9}$ | < 300 | $< 1\ 10^{-5}$ | < 20 |
| Neoprene connector boot | 1400±400 | $(1.1±0.3)\ 10^{-7}$ | 1300±300 | $(3.1±0.7)\ 10^{-8}$ | 1600±250 | $(5.1±0.8)\ 10^{-5}$ | < 18 |
| Jupiter connector O-rings | 1400±340 | $(1.1±0.3)\ 10^{-7}$ | 1200±370 | $(2.9±0.9)\ 10^{-7}$ | 7900±300 | $(2.5±1.1)\ 10^{-4}$ | |
| **Optical fiber** | | | | | | | |
| Teflon for optical fibers coating | 28±4 | $(2.3±0.3)\ 10^{-9}$ | < 6 | $< 1.4\ 10^{-9}$ | < 60 | $< 2\ 10^{-6}$ | < 2.4 |
| Kevlar | 430±80 | $(3.5±0.6)\ 10^{-8}$ | <140 | $< 3.4\ 10^{-8}$ | < 870 | $< 2.8\ 10^{-5}$ | < 68 |
| EPO-TEK 353ND resin | < 47 | $< 4\ 10^{-9}$ | < 24 | $< 6\ 10^{-9}$ | < 320 | $< 1\ 10^{-5}$ | < 17 |
| **Stainless steel samples** | | | | | | | |
| AISI304L for SSS | 4.6±0.9 | $(3.7±0.7)\ 10^{-10}$ | 11.4±1.1 | $(2.8±0.3)\ 10^{-9}$ | < 14 | $< 4.5\ 10^{-7}$ | 6±1 |
| Piping steel | < 14 | $< 1.1\ 10^{-9}$ | < 10 | $< 2.5\ 10^{-9}$ | < 34 | $< 1.1\ 10^{-6}$ | 14±4 |
| Steel for flanges | 6.2±1.2 | $(5±1)\ 10^{-10}$ | 6.5±1.6 | $(1.6±0.4)\ 10^{-9}$ | < 13 | $< 4.2\ 10^{-7}$ | 14±1 |
| Steel for storage vessel | 17±3 | $(1.4±0.2)\ 10^{-9}$ | 3.8±2.6 | $(9.3±6.4)\ 10^{-10}$ | < 19 | $< 6\ 10^{-7}$ | 12±2 |
| Steel TK3B for storage vessel | 5±1 | $(4±1)\ 10^{-10}$ | 5±2 | $(1.2±0.5)\ 10^{-9}$ | 4±2 | $(1.3±0.6)\ 10^{-7}$ | 46±3 |
| Steel foil for Rn emanation test | 0.6±0.2 | $(5±2)\ 10^{-11}$ | 0.2±0.1 | $(5±3)\ 10^{-11}$ | 1.8±0.6 | $(6±2)\ 10^{-8}$ | 18±1 |
| **Miscellaneous** | | | | | | | |
| Nylon Curbell 5, 6" rod for caps CTF | < 0.6 | $< 0.5\ 10^{-11}$ | < 0.5 | $< 1.2\ 10^{-10}$ | < 3 | $< 1\ 10^{-7}$ | < 0.2 |
| Nylon 2.5" Pipe for IV connection | < 1.2 | $< 9.3\ 10^{-11}$ | < 0.9 | $< 2.2\ 10^{-10}$ | < 18 | $< 6\ 10^{-7}$ | < 0.4 |
| Nylon fishing line, monofilament. | < 1.1 | $9\ 10^{-11}$ | < 0.6 | $< 1.5\ 10^{-10}$ | 4.8±2.5 | $(1.6±0.8)\ 10^{-7}$ | < 0.4 |
| Tensylon rope for IV | < 0.3 | $< 2.4\ 10^{-11}$ | < 0.3 | $< 7\ 10^{-11}$ | 15±2 | $(4.7±0.5)\ 10^{-7}$ | < 0.1 |
| SilicaGel 60 for purification column | 100±50 | $(8±4)\ 10^{-9}$ | < 160 | $< 4\ 10^{-8}$ | < 600 | $< 2\ 10^{-5}$ | < 90 |
| Charcoal for radon adsorption | < 300 | $< 2\ 10^{-8}$ | < 500 | $< 1\ 10^{-7}$ | < 2000 | $< 6\ 10^{-5}$ | < 100 |

The overall activity of the PMT assembly is mainly caused by the glass. The sand that is used in the glass production is purer than the end product, therefore the other additives or procedures must cause the higher contamination. The base glass (holding the electrical connection pins) has a larger distance to the inner vessel than the rest of the PMTs, so that the higher contamination in U/Th is tolerable, but the much higher K concentration is noticeable. Another measured batch of that type of glass showed a K concentration about 100 times less, but slightly higher $^{226}$Ra and $^{228}$Th concentrations.

The sensitivity that has been achieved in each measurement depends not only on the efficiency and the background of the spectrometer in use, but also on the measuring time and the amount (mass) of the sample. This is clearly seen in the results of the stainless steel measurements. The lowest concentration for U, $^{226}$Ra and Th was obtained for a 38.1 kg sample (steel foil) measured with GeMPI for 21 days. On the low sensititvity side was the sample 'piping steel' having a weight of about 1 kg and a measuring time of about 2 days. As can be seen in Table 3, the contamination of stainless steel by primordial radionuclides can vary by almost two orders of magnitude.

## 4 Ultratrace determinations of Radon

One of the most dangerous contaminants for low energy experiments and for BOREXINO especially is $^{222}$Rn and its daughter nuclides. It is omnipresent in the environment in amounts that have concentrations up to 7 orders of magnitude above what is tolerable in the BOREXINO detector. The chief sources of $^{222}$Rn are the terrestial $^{222}$Rn carried by the laboratory air and $^{226}$Ra (its progenitor) present as contamination in detector materials.

All of the construction materials which are close to the scintillator or that are in contact with it must be examined for $^{222}$Rn emanation at a sensitivity down to the 100 µBq level and for $^{222}$Rn permeability down to $10^{-12}$ cm$^2$/s [10]. Furthermore, the air of all rooms in which these materials are handled has to be monitored for $^{222}$Rn at a level of 1 mBq/m$^3$. As the last step in the purification procedure used for the scintillator of BOREXINO $^{222}$Rn, $^{85}$Kr and other dissolved gases are removed by stripping using gaseous N$_2$. The $^{222}$Rn contamination of the N$_2$ for this purpose should not exceed the µBq/m$^3$ range. Finally the shielding water in the CTF has to be assayed for $^{222}$Rn and $^{226}$Ra contents before the filling and during the operation, to assure the required purity at the mBq/m$^3$ level.

### 4.1 Radon sample preparation and counting

All three assay procedures (emanation, nitrogen monitoring, and water monitoring) used in the BOREXINO project utilize $^{222}$Rn collection on charcoal and internal proportional counting. The needed levels of high sensitivity require stringent control of $^{222}$Rn contributions from $^{226}$Ra impurities and atmospheric $^{222}$Rn in the analytic procedures. Generally, this contribution to the signal is measured using a "blank run" and is denoted the "blank signal". To keep the $^{222}$Rn levels low, a special charcoal with only 0.3 mBq/kg $^{226}$Ra is used. In addition metal sealed high vacuum techniques are used. Before counting, the sample needs to be cleaned from gaseous impurities. This is performed in the vacuum glass purification systems (one at MPI in Heidelberg and one at LNGS) by the help of traps with various absorber materials. The traps are operated at different temperatures in such a way that either the impurities are trapped and the $^{222}$Rn allowed to pass or *vice versa*. As a final step in the purification process, the remaining chemically reactive gases are adsorbed on an Al/Zr getter.

The $^{222}$Rn sample is then mixed with the counting gas (P10, 90% Ar + 10% CH$_4$) and filled into proportional counters [11-13] which have an active volume of about 1 cm$^3$ and a filling pressure of about 1 bar. Under these conditions the large energy deposition of the three α particles of $^{222}$Rn and its short lived daughter nuclides (≥ 50 keV) allow an effective discrimination against the electron and photon dominated background. The total counting efficiency corresponds to 1.24 and 1.48 counts per $^{222}$Rn-decay for the two different types of counters in use. The background above 50 keV ranges from 0.1 to 2 counts per day, depending on the minor variations associated with individual counters. However, after extensive measurements of high activities this background increases due to the accumulation of $^{210}$Pb, a long-lived radionuclide ($T_{1/2}$ = 22.3 y), on the walls of the outer cylindrical cathode. Its α-emitting daughter nuclide $^{210}$Po can deposit energies above 50 keV in the counter.

The sample preparation procedure contributes an amount of $^{222}$Rn to the signal in the range of 10 to 50 µBq, the uncertainty is around 20 µBq. With these numbers, the resulting sensitivity ranges down to about 50 µBq.

### 4.2 The emanation apparatus

The emanation apparatus consists, as sketched in Figure 2, of two electropolished stainless steel chambers of 20 l and 80 l volume with large metal sealed top flanges for sample insertion. The chambers are each equipped with spargers to allow for an effective removal of $^{222}$Rn from any liquid samples. All connections are made using electropolished stainless steel tubing, and all flanges are metal sealed. After sample insertion, residual $^{222}$Rn is removed either by applying vacuum or, if the sample cannot be evacuated, by flushing with $^{222}$Rn free He (passed over charcoal at liquid nitrogen temperature).

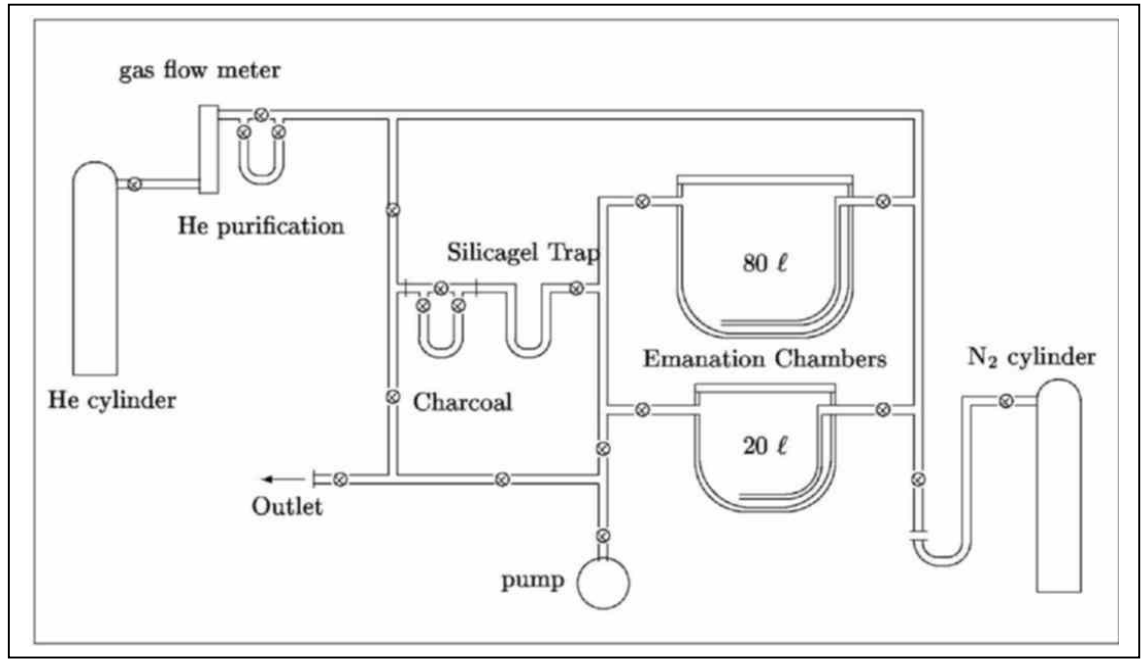

Figure 2: Sketch of the emanation apparatus

After several days of ingrowth of $^{222}$Rn from $^{226}$Ra, the chamber is then flushed with He, the $^{222}$Rn extracted and trapped in a final step on charcoal. A silica gel trap before the charcoal removes water vapor in case of wet samples. Contributions to the signal from sources other than the investigated sample (blank signal) correspond to (64 ± 15) µBq for the small chamber and (255 ± 30) µBq for the big one, allowing a sensitivity level as low as 70 µBq and 100 µBq, respectively. For counting the $^{222}$Rn from small samples, glass vials of different volumes (up to 1000 cm$^3$) can be directly connected to the preparation lines. They give no measurable background contribution to the signal [13,14]. A large variety of samples have been examined for $^{222}$Rn emanation. Table 4 lists a selection of these measurements that demonstrate the performance of the system, or are of special importance for the experiment. The quoted uncertainties correspond to 1σ; the limits are given at a 90% confidence level.

Table 4. Emanation of $^{222}$Rn from materials used in the BOREXINO experiment.

| Sample | Surface/Amount | Activity (mBq) | Vessel |
|---|---|---|---|
| PMT | 1 piece | < 0.031 | Small chamber |
| Master Bond ZP45HT | 340 cm$^2$ | < 0.86 /m$^2$ | Small chamber |
| µ-metal sealing | 0.06 m$^2$ | < 0.43 /m$^2$ | Big chamber |
| TYVEK 1073B | 16 m$^2$ | < 0.0039 /m$^2$ | Big chamber |
| Nitril | 96 cm$^2$ | (187 ± 9) /m$^2$ | Glass vial |
| PERMATEX | 280 cm$^2$ | (24.9 ± 4.2) /m$^2$ | Glass vial |
| Safety valve | 1 piece | 0.08 ± 0.03 | Direct |
| BUSTO | 139 cm$^2$ | (1750 ± 40) /m$^2$ | Glass vial |
| Butyl-O-rings | 0.66 m$^2$ | (19.6 ± 0.5) /m$^2$ | Small chamber |
| Polyurethan 5 flat gaskets | 0.27 m$^2$ | (0.3 ± 0.1) /m$^2$ | Small chamber |
| Steel foil | 71 m$^2$ | (0.0102 ± 0.0006) /m$^2$ | Big chamber |
| Steel foil (after water rinsing) | 71 m$^2$ | (0.0046 ± 0.0010) /m$^2$ | Big chamber |

The first three samples are materials which are installed in the inner detector: a photo multiplier tube (PMT), an epoxy resin for the sealing of the PMTs and a second resin, which covers the µ-metal shielding. For all of these samples only upper limits are quoted, as is the case for most of the measured materials of the inner detector. Scaled to the total detector, the emanation rate of these parts will be tolerable in BOREXINO (taking into account the additional $^{222}$Rn barrier foil, see Figure 1).

The inner part of the detector is enclosed in a stainless steel sphere and all the PMTs are encapsulated in steel cans. It is important then to measure the $^{222}$Rn emanation from steel. To estimate that, a sample of known $^{226}$Ra contamination (see Table 3) was measured: the result suggests that most of the $^{222}$Rn is produced from surface contamination. This discovery is further supported by comparison to a follow up measurement after a simple water rinse. For solid materials, such as steel, one has to assume that $^{222}$Rn atoms can escape from the surface only by the recoil received by the decay of $^{226}$Ra. The expected emanation rate from the $^{226}$Ra measured by Ge-spectrometry (Table 3) is more than one order of magnitude lower than what is measured (after rinsing).

The TYVEK (tradename) material is a white foil that will cover the inner surface of the outer tank for diffuse reflection of the Čerenkov light produced by muons. Nitril is a black rubber used for the sealing of the 4 m × 4 m door of the outer tank. PERMATEX (tradename) is an epoxy resin that covers the inner walls of the CTF tank. PERMATEX is known to contain $^{226}$Ra: the emanation from this material has been most probably the major source of $^{222}$Rn in the CTF shielding water during the first measurement period [4]. A larger number of safety valves like that reported in Table 4 are installed in the nitrogen supply plant: their contributions to the $^{222}$Rn contamination is consequently only very minor. BUSTO (tradename), a standard sealing material that was originally used in some flanges of the nitrogen plant, had to be replaced by Butyl because of its very high emanation rate of $^{222}$Rn. Polyurethan was found in most cases to be even lower in $^{222}$Rn emanation than Butyl, despite its slightly higher permeability for $^{222}$Rn.

### 4.3 The water Ra/Rn assay system

The main part of the water assay system (Figure 3) is a 480 l electropolished stainless steel tank that can be pressurized up to 3 bars. It is equipped with a sparging system, a vacuum pump, integrating flowmeters (for both the incoming water and the sparging gas), and indicators for pressure, temperature and water level. The system operates by sparging the $^{222}$Rn from the water, concentrating it and then loading it into miniaturized low background proportional counters. The line to concentrate the $^{222}$Rn from the stripping gas is directly mounted on the tank; it consists of a cooling trap filled with copper wool and a charcoal trap. The water tank has a large top flange with an elastomer (polyurethane) o-ring and an additional seal made of indium. All other flanges are metal sealed.

An operation cycle starts with the tank evacuation to remove the residual $^{222}$Rn and to allow water filling (sample size about 350 l). The filling is made without venting in order to avoid $^{222}$Rn in-leakage as well as uncontrolled $^{222}$Rn loss. The tank is next immediately flushed with ~ 5 m$^3$ of high purity nitrogen from our unique underground nitrogen plant (see section 4.4). The water vapor in the effluent stream of the nitrogen is then adsorbed in the trap filled with copper wool at -55 $^o$C. Finally, the $^{222}$Rn is collected on the charcoal trap at -196 $^o$C. The entire extraction procedure lasts about two hours.

The $^{226}$Ra concentration of the same water sample is measured by means of a second extraction following the same procedure as for the first $^{222}$Rn determination, only this time a few days later in order to count the $^{222}$Rn from the $^{226}$Ra which has been accumulated during the ingrowth period. The blank value (at saturation) of (1.5 ± 0.1) mBq results in a detection limit of about 1 mBq/m$^3$ for $^{226}$Ra. In the $^{222}$Rn measurement the residence time of the water in the tank is much shorter than for the $^{226}$Ra analysis, thus the applicable blank for the actual $^{222}$Rn measurement is lower. Hence, the actual sensitivity for this measurement is about 0.1 mBq/m$^3$. It is limited mainly by the blank activity of the follow-on sample preparation.

The efficiency of the $^{222}$Rn recovery from water was determined by adding a small amount of water with a relatively high and well-known $^{222}$Rn activity to the tank prior to the extraction. The recovery yield of (97 ± 4) % is consistent with the expected 100%. A successive second extraction in this test showed that less than 0.5% of the activity remained in the tank [15]. A sketch of the entire system is displayed in Figure 3.

In Table 5 the results of some measurements of the shielding water, as purified under different conditions by the water purification plant in BOREXINO, are shown. There are two units which reduce the $^{222}$Rn concentration by nitrogen sparging, a buffer tank of 5 m$^3$ volume and the (more effective) stripping column of 6 m height. As can be seen, the $^{222}$Rn concentration decreases with increasing nitrogen flow, while it increases along with the water flow. In the circulation mode, the $^{222}$Rn concentration can be reduced to the level of some mBq/m$^3$. The measurements demonstrate that the required sensitivity for $^{226}$Ra and $^{222}$Rn has more than been reached and that the water plant can produce water that is pure enough at the design level.

In the CTF water we have measured similar low $^{226}$Ra concentrations, whereas $^{222}$Rn is slightly higher than one would expect from the $^{222}$Rn decay of the original water filling. There is indication of a source at the tank wall (emanation from the PERMATEX wall liner or by $^{226}$Ra from water absorbed in the tank wall), which is supported by a slightly higher $^{222}$Rn concentration (about 30 mBq/m$^3$) at the wall.

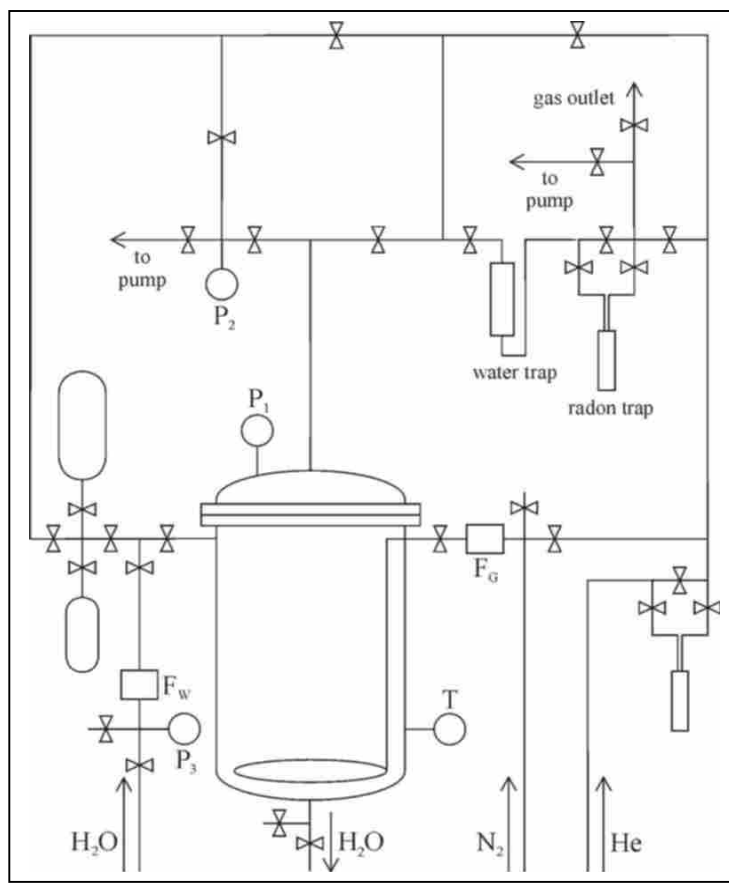

Figure 3: Sketch of the water $^{226}$Ra/$^{222}$Rn assay system. $F_W$ and $F_G$ are flow meters for water and gas, T indicates the temperature sensor, the pressure indicators are labeled with $P_1$, $P_2$ and $P_3$. The two vials on the left hand side are used for the determination of the $^{222}$Rn recovery efficiency.

Table 5: $^{222}$Rn/$^{226}$Ra measurements of water purified by the BOREXINO water purification plant. Samples have been taken under different working conditions.

| Sample | Production conditions | | | $^{222}$Rn activity [mBq/m$^3$] | $^{226}$Ra activity [mBq/m$^3$] |
| --- | --- | --- | --- | --- | --- |
| | H$_2$O flow [m$^3$/h] | Buffer tank N$_2$ flow | Stripping column N$_2$ flow [kg/h] | | |
| 1 | 2 | low | ~ 8 | 13040 ± 360 | --- |
| 2 | 2 | high | 30 | 704 ± 7 | 1.2 ± 0.5 |
| 3 | 1 | medium | 30 | 483 ± 10 | 0.9 ± 0.4 |
| 4 | 1 | high | 20 | 247 ± 6 | 3.8 ± 0.7 |
| 5 | 1 | high | 30 | 186 ± 5 | 2.0 ± 0.6 |
| 6 | circulation mode: multiple deionization and sparging | | | 3.0 ± 0.4 | 1.3 ± 0.9 |

## 4.4 Concentration line for radon in nitrogen

In BOREXINO we have developed an extremely low level ($^{222}$Rn) nitrogen supply plant. This plant incorporates a special liquid nitrogen purification column using a charcoal bed low in $^{226}$Ra. The purification advantage of using the liquid versus the gaseous phase is that the volume of the medium to be purified is smaller than it would be had the nitrogen been a gas, thus the opposing entropy term (reflecting the tendency of the material to oppose being on the column) is reduced in magnitude.

To obtain a sufficiently strong signal in the testing of this plant, $^{222}$Rn is collected from the nitrogen gas by a concentration line. This line allows us to process several hundred m$^3$ of gas during the sampling process. Two charcoal columns operated at -196 $^o$C, each containing 1.2 l of charcoal, can be used alternatively or in series. Prior to the charcoal column, the nitrogen passes over a particle filter made from TEFLON (tradename). At the exit, the total amount of nitrogen is monitored through a flow meter.

Following the collection step, the charcoal trap is warmed up and the sampled $^{222}$Rn is transferred using $^{222}$Rn free helium to a smaller removable charcoal trap. This trap is used to transfer the sample to the $^{222}$Rn glass purification and handling system (see section 4.1). The concentration line includes some additional components in order to handle the gases. These include vacuum pumps and pressure indicators.

The $^{222}$Rn emanation from the charcoal traps was measured to be (60 ± 20) µBq and (117 ± 52) µBq, respectively (in equilibrium, after a large number of $^{222}$Rn half-lives). The absolute contribution of blank (background) $^{222}$Rn to the signal depends on the duration of the concentration procedure. It is determined by making a measurement using the maximum gas flow through the concentration line of about 20 m$^3$/h. For typical samples having total flow through volumes in the range of 100 m$^3$ the contribution is (7 ± 2) µBq, while for a much larger but still typical 500 m$^3$ sample it is (27 ± 9) µBq. As a consequence the sensitivity in our measurements is better than 0.5 µBq/m$^3$. This represents a significant jump in our capability to manage few numbers of radioactive rare gas atoms (here $^{222}$Rn) during the processing of gas volumes at hundreds of cubic meters and represents the state of the art in the field.

The capacity of the charcoal traps has been tested to be large enough for samples in excess of 500 m$^3$ by using both charcoal traps in series: the activity collected in the second trap was consistent with zero. The recovery of $^{222}$Rn from the traps has been tested to be (100 ± 5) % by introducing a calibrated activity into a small sample (30 m$^3$) [16,17].

The nitrogen supply plant for BOREXINO produces two different qualities of gaseous nitrogen with respect to the $^{222}$Rn contamination. The so-called standard purity nitrogen is produced by evaporating liquid nitrogen with normal atmospheric evaporators. High purity nitrogen is purified in the liquid phase and evaporated with a special small surface evaporator [16].

Table 6 shows the results of some measurements of standard and high purity nitrogen. They clearly demonstrate that the most challenging sensitivity (see Table 1) for $^{222}$Rn in high purity nitrogen (used for the scintillator sparging) was met. $^{222}$Rn concentrations in unpurified nitrogen have been determined to be in the range of 50 µBq/m$^3$.

Table 6: $^{222}$Rn concentration in the nitrogen provided by the nitrogen plant of BOREXINO.

| Sample size (m$^3$) | Gas flow (m$^3$/h) | Quality | Activity (µBq/m$^3$) |
| --- | --- | --- | --- |
| 30 | 19 | Regular | 71 ± 9 |
| 35 | 50 | Regular | 27 ± 4 |
| 40 | 100 | Regular | 51 ± 7 |
| 120 | 50 | High purity | 0.48 ± 0.32 |
| 500 | 100 | High purity | 0.78 ± 0.10 |
| 500 | 100 | High purity | 0.30 ± 0.09 |

The spread in the $^{222}$Rn concentration in nitrogen of the same quality is not yet completely understood; it might reflect a variable contamination of the supplied liquid nitrogen or different emanation rates in the three liquid nitrogen storage tanks, possibly also depending on their filling level.

The system for $^{222}$Rn assay described above has been extensively used also for subsystems of the BOREXINO experiment, such as the large columns of the liquid handling system and the storage vessels. The primary interest is to test for $^{226}$Ra purity and tightness against airborne $^{222}$Rn. For this purpose the accumulated $^{222}$Rn emanation in these units is flushed with high purity nitrogen into the traps of the assay system. There are plans to use this system to measure also other radioactive noble gases, such as $^{39}$Ar and $^{85}$Kr.

## 4.5 Radon monitoring in air

In order to monitor $^{222}$Rn in air at a level of 1 mBq/m$^3$ a detector based on the method of electrostatic collection of the positively charged daughter ions $^{218}$Po and $^{214}$Po onto an α detector has been built for BOREXINO [18]. The measured α activity of $^{218}$Po and $^{214}$Po is correlated with the $^{222}$Rn concentration in the air. The detector has a volume of 418 l and operates at a voltage of 30 kV. The sensitivity of this type of detector depends on the collection efficiency which in turn is influenced by the degree of contamination of the air with neutralizing chemical reagents such as water or hydrocarbons. The value which has been reached in measurements for BOREXINO is (27 ± 0.8) counts/day per mBq/m$^3$ of $^{222}$Rn if the count rates for both $^{218}$Po and $^{214}$Po are summed up. Since the background count rate in the region of interest was measured to be less than 2 counts/day, this results in a detection limit for $^{222}$Rn of less than 1 mBq/m$^3$.

# 5 Neutron activation analysis

## 5.1 Basic principles

The most common reaction for the determination of trace elements by Neutron Activation Analysis (NAA) is $^{A}Z + n \Rightarrow\ ^{A+1}Z + \gamma$. The nucleus $^{A+1}Z$ decays with a certain half life $T_{1/2}$ by $\beta^-$ or $\beta^+$ emission or electron capture to excited levels of the corresponding daughter nucleus. This daughter nucleus then in turn is de-excited by one or more characteristic γ-rays. These γ-rays are detected in a NaI- or Ge-spectrometer, thereby allowing the determination of the concentration of a nucleus $^{A}Z$ in a matrix by measuring the activity of the nucleus $^{A+1}Z$. The activity of $^{A+1}Z$ is governed by the following parameters: neutron capture cross section, neutron flux, exposure time and decay time. Since not only the nuclei of the trace element Z but also the matrix is exposed to the neutrons, it helps if the activity produced by the matrix is less or at most equal to the activity of $^{A+1}Z$ during the time of the measurement in the relevant energy range.

The interference from the neutron produced matrix activity usually restricts the application of NAA to matrices which have much lower neutron capture cross sections and/or much shorter half lives than the isotopes to be detected. These demands are satisfied by light elements (e.g. H, C, N, O, F and Si) which are often constituents of matrices. In other cases, the matrix problem can be reduced by a chemical separation of the trace elements from the matrix or by the use of long cooling times if the matrix activity decays faster than the activity of the trace element. This makes NAA an ideal tool for the detection of trace constituents in $H_2O$, Si, organic matrices and plastics, which are used for the construction of low-level detectors.

In addition, the combination of neutron activation and low-level counting methods opens the possibility to analyze the activity of the parent isotopes $^{238}U$ and $^{232}Th$ in two main components of BOREXINO and the CTF, the shielding water and the scintillator, at a level comparable to the activity of their daughter nuclides $^{226}Ra$ and $^{228}Th$.

The neutron activations in this work have been carried out at the research reactor FRMI in Garching, Germany. 250 g samples of water or liquid scintillator were irradiated for typically 100 hours at a neutron flux between $10^{12}$ $s^{-1}cm^{-2}$ and $10^{13}$ $s^{-1}cm^{-2}$. A special irradiation facility has been developed at the reactor for liquid samples subject to radiolysis: the gas produced during the irradiation is released into the pool water to avoid the buildup of high pressures inside the sample container.

Currently, the FRMI reactor has been shutdown and the completely new FRMII reactor is near completion. We expect to continue the measurements described in this section as soon as the new reactor is operating.

## 5.2 Increasing the sensitivity of NAA

The sensitivity of NAA is not only limited by the matrix but it is also determined by the activation parameters. The samples have to be prepared in clean boxes or clean rooms and handled in PFA (perfluoroalkoxy polymeric materials) and quartz vessels before the irradiation and in synthetic quartz vessels during the irradiation, in order to reduce the blank value of the analysis [19]. The irradiation parameters of sample mass and neutron flux should be as large as possible while the irradiation time can be adjusted to the matrix and the trace elements to be determined. In addition it is necessary to keep the activity in the counting system from the irradiated sample or from the counting system itself as low as possible.

The radiochemical separation of the activated nuclei from the matrix can improve the sensitivity significantly, as discussed above. In the ideal case the induced activity can be split into two or more groups containing the matrix activity, the activity of other interfering traces and the one of the radionuclides of interest. Because of the severe requirements of solar neutrino experiments we have pursued many techniques in this regard.

Another important factor which governs the sensitivity is the detection efficiency of the emitted γ-rays. The signal-to-noise ratio cannot be improved above a certain limit by the increase of efficiency due to physical and financial restrictions; however, an improvement can be achieved if coincidence or anticoincidence counting modes are applied or even more simply, if all the background is removed through the use of low background counting. This can be realized for instance by γ-γ-(anti)coincidence or β-γ-(anti)coincidence counting modes. For some isotopes even delayed β-γ-e$^-$-coincidences may be applied if the indicator radionuclide populates an isomeric state providing an additional signature of the decay [20,21].

### 5.3 Experimental set up for β-γ coincidences

In order to measure the emitted particles as well as the photons, the radiochemically processed sample is dissolved in a liquid scintillator which is viewed from one side by a photomultiplier and from the other side by a Ge-detector. The two detectors provide both timing and energy information which allow the trigger conditions tuned to the decay of the indicator radionuclide. The set-up we have used in connection with different types of Ge-detectors is shown in Figure 4. The information from the two detectors is recorded by a CAMAC data acquisition system. The recorded data can be analyzed off-line. Events are selected by their energy (γ-rays, β-particles and conversion electrons) and the time difference between β decay and conversion electron emission.

Currently our main interest has been in the U and Th determinations. Using our techniques we have set records and we expect that we are still able to improve upon these numbers. Below we describe some of the measurements.

For the case of U we have developed a number of techniques. Besides the delayed β-γ-conversion electron coincidence used for the determination of the U (see below), we have applied the β-γ-coincidence counting mode to determine Th and Lu in different samples. Although the β-γ-coincidence does not reduce the intrinsic background due to a similar β endpoint energy of Th and Lu compared to the interfering nuclides, the external background originating from the activity in the counting chamber can be suppressed. This β-γ-coincidence can be applied effectively if the intrinsic activity is low; the sample has to be counted for several weeks, with the danger that the external background would worsen the detection limit for indicator radionuclides with half lives of weeks or months.

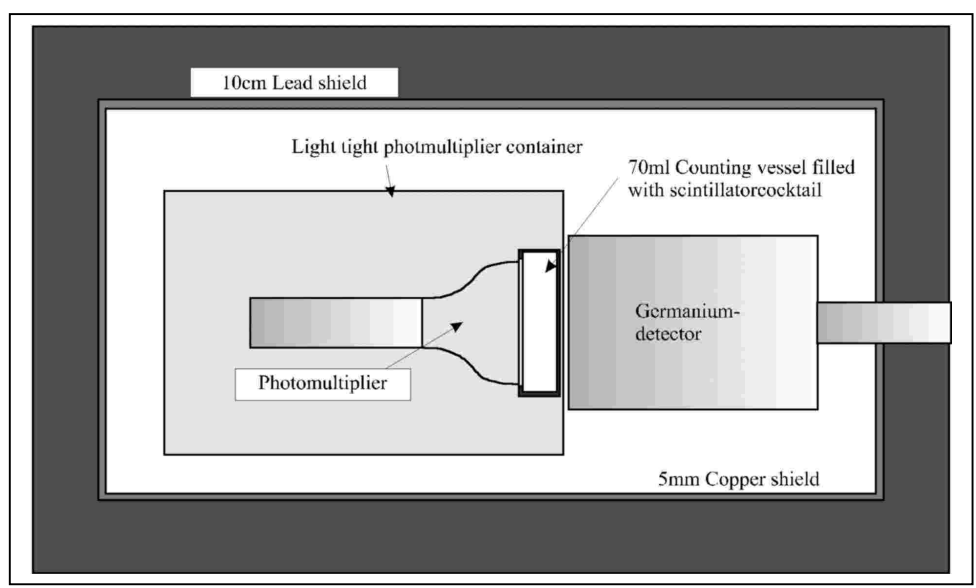

Figure 4: Experimental set-up used for the measurement of a sample dissolved in a liquid scintillator

An effective background suppression can be achieved by a β-γ-coincidence if the β-endpoint of the indicator radionuclide differs significantly from the β-endpoint of the interfering nuclides. This allows us to set a "cut" on the β-endpoint energy to suppress the interfering activity. Such β-γ anticoincidences can be used as well to search for radionuclides decaying via electron capture such as $^{51}$Cr, $^{64}$Cu, $^{65}$Zn, $^{75}$Se, $^{85}$Sr or $^{97}$Rb if the interfering activity is associated to the β-γ emission. Since the intrinsic sample activity is often caused by pure β emitters, it can be suppressed effectively by cuts on the β endpoint energy, on the conversion electron energy and on the delay time between emitted β particle and conversion electrons.

The indicator nuclide for $^{238}$U is $^{239}$Np which is produced via neutron capture and the subsequent β$^-$ decay of $^{239}$U. With a probability of 40.5% the β$^-$ decay of $^{239}$Np in turn populates an isomeric state in $^{239}$Pu which decays through the emission of γ-rays and conversion electrons. This implies that the number of measured conversion electrons decays exponentially with the half life (193 ns) of that isomeric state, offering the possibility not only to use the information on the energy of the emitted β particles, γ-rays and conversion electrons, but also the time delay between β particles and conversion electrons as a signature to discriminate interfering activities. The optimization of all these factors can improve the sensitivity by some orders of magnitude. This reveals the capability of coincidence measurement techniques in combination with radiochemical neutron activation analysis (RNAA) to suppress interfering radionuclides as compared to instrumental neutron activation analysis (INAA) [22]. Figure 5 shows the γ spectrum of an irradiated liquid scintillator sample prior to and after a radiochemical separation, combined with coincidence counting techniques. In the latter only the γ-ray peaks from $^{239}$Np around 100 keV survive the cuts. The cut efficiency can be checked by applying the same cuts to a known activity of $^{239}$Np.

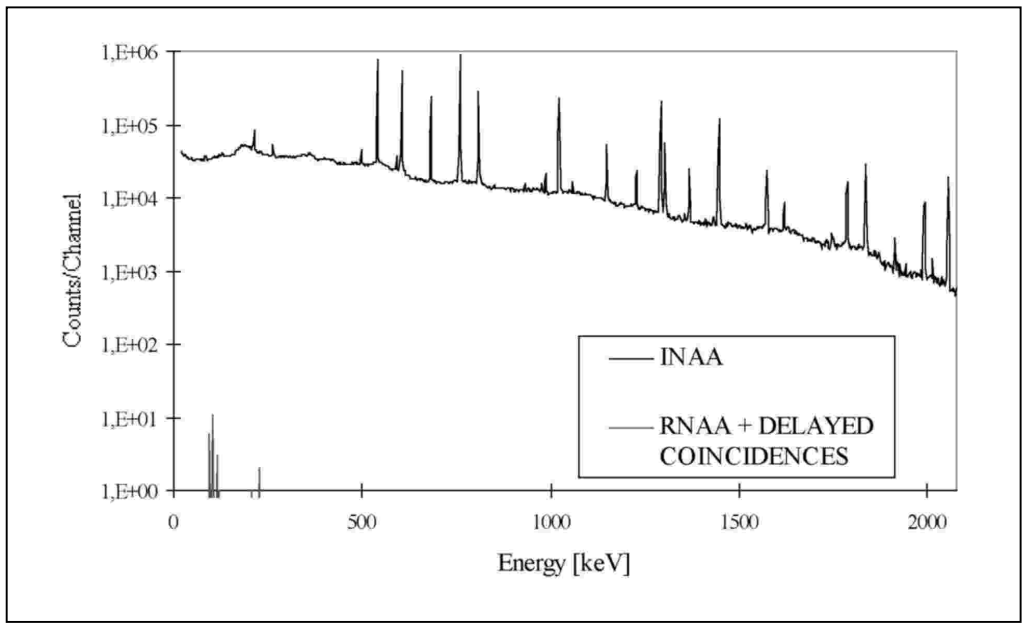

Figure 5: The γ spectrum of an irradiated liquid scintillator prior to and after a radiochemical separation, combined with coincidence counting techniques. The coincidence counting mode was adjusted to the detection of $^{239}$Np which is the indicator radionuclide for $^{238}$U (see text).

## 5.4 Results: trace elements in water

The analysis of water is relatively simple as the water can be evaporated after irradiation and the evaporation residue can be processed radiochemically without any pre-treatment. The radiochemical treatment divides the evaporation residue into two groups [23]. One group contains the indicator radionuclides with short half lives (of the order of days); their decays are measured directly. The second group contains Np (U), Pa (Th), Lu and is dissolved in a liquid scintillator cocktail, which detects the β-γ-conversion electrons and the β-γ-coincidences, as described above. The results obtained for water are shown in Table 7.

Table 7: Trace element concentration in subboiled high purity water [24]. Uncertainties (1σ) of positive numbers are about 50%, limits are given for a 90% confidence level.

| Element | Concentration [g/g] | Element | Concentration [g/g] |
|---|---|---|---|
| Au | $< 5 \cdot 10^{-15}$ | Lu | $< 7 \cdot 10^{-14}$ |
| Cd | $< 4 \cdot 10^{-13}$ | Na | $< 1 \cdot 10^{-13}$ |
| Co | $3 \cdot 10^{-13}$ | Rb | $< 9 \cdot 10^{-13}$ |
| Cr | $< 4 \cdot 10^{-12}$ | Sb | $< 7 \cdot 10^{-14}$ |
| Fe | $< 3 \cdot 10^{-11}$ | Sc | $< 6 \cdot 10^{-14}$ |
| In | $< 5 \cdot 10^{-13}$ | Th | $2 \cdot 10^{-15}$ |
| Ir | $< 5 \cdot 10^{-15}$ | U | $4 \cdot 10^{-15}$ |
| K | $< 8 \cdot 10^{-13}$ | W | $< 2 \cdot 10^{-14}$ |
| La | $< 4 \cdot 10^{-15}$ | Zn | $< 2 \cdot 10^{-12}$ |

## 5.5 Results: trace elements in organic scintillators

We have investigated the two scintillators which were tested in the CTF (see section 8). The first one is a scintillator consisting of 1,2,4-Trimethylbenzene (PC) and 2,5-Diphenyloxazole (PPO). The PPO is dissolved as a primary wavelength shifter at a concentration of 1.5 g/l in the solvent. The second one is composed of Phenyl-o-xylylethane (PXE) and p-Diphenylbenzene (para-terphenyl, pTP) which is added as a primary wavelength shifter at a concentration of 2 g/l to the solvent along with a secondary wavelength shifter 1,4-bis (2-methylstyryl) benzene (bis-MSB) at the 20 mg/l level.

Although organic solvents are severely degraded and decomposed by the neutron irradiation, they can be irradiated up to 100 hours in the irradiation facility with appropriate gas release [25]. The radiochemical process following the irradiation, after a cooling time of two or three days, is more complicated than in the case of aqueous samples. The irradiated solvents are subjected to an acidic extraction, which transfers the activated trace elements into an acid. This is later evaporated and split up into two groups by a liquid-liquid extraction or by ion exchange, containing activities which can be measured by the counting modes mentioned above [23]. The results of some trace elements in scintillator samples are given in Tables 8 and 9.

The specific activity for the detection limits shows that the analytical procedure presented is capable to set limits on the detector impurities which corresponds to a few decays per kg and year. This sensitivity can not only be used for quality assurance of the materials but also for the development of cleaning and refining procedures.

Table 8: PC/PPO based scintillator (see text). Uncertainties (1σ) for positive numbers are about 50%, limits are given for a 90% confidence level.

| Element | Concentration [g/g] | Element | Concentration [g/g] |
|---------|---------------------|---------|---------------------|
| Ag | $4 \cdot 10^{-12}$ | Lu | $< 8 \cdot 10^{-16}$ |
| Au | $2 \cdot 10^{-14}$ | Mo | $< 5 \cdot 10^{-13}$ |
| Cd | $< 1 \cdot 10^{-12}$ | Na | $< 2 \cdot 10^{-12}$ |
| Ce | $< 2 \cdot 10^{-13}$ | Rb | $< 7 \cdot 10^{-13}$ |
| Co | $3 \cdot 10^{-14}$ | Sb | $3 \cdot 10^{-14}$ |
| Cr | $7 \cdot 10^{-13}$ | Sc | $< 3 \cdot 10^{-15}$ |
| Cs | $< 1 \cdot 10^{-13}$ | Sm | $< 2 \cdot 10^{-15}$ |
| Fe | $< 2 \cdot 10^{-11}$ | Th | $< 2 \cdot 10^{-15}$ |
| In | $< 2 \cdot 10^{-12}$ | U | $< 2 \cdot 10^{-16}$ |
| Ir | $< 2 \cdot 10^{-14}$ | W | $< 3 \cdot 10^{-14}$ |
| K | $< 4 \cdot 10^{-12}$ | Zn | $< 4 \cdot 10^{-12}$ |
| La | $2 \cdot 10^{-14}$ | | |

Table 9: PXE/pTP based scintillator (see text). The limits are given for a 90% confidence level.

| Element | [g/g] | Element | [g/g] |
|---------|-------|---------|-------|
| Ag | $(2.3 \pm 0.4) \cdot 10^{-13}$ | Lu | $< 3.8 \cdot 10^{-16}$ |
| Au | $< 3.8 \cdot 10^{-17}$ | Mn | $< 2.4 \cdot 10^{-9}$ |
| Ba | $< 8.3 \cdot 10^{-13}$ | Mo | $< 3.7 \cdot 10^{-14}$ |
| Cd | $< 8.3 \cdot 10^{-15}$ | Rb | $< 1.1 \cdot 10^{-13}$ |
| Ce | $< 2.1 \cdot 10^{-14}$ | Re | $< 5.5 \cdot 10^{-16}$ |
| Co | $(2.6 \pm 0.4) \cdot 10^{-14}$ | Sb | $(8.7 \pm 1.3) \cdot 10^{-15}$ |
| Cr | $(2.3 \pm 0.4) \cdot 10^{-13}$ | Sc | $< 2.3 \cdot 10^{-16}$ |
| Cs | $< 2.1 \cdot 10^{-13}$ | Sn | $< 3.3 \cdot 10^{-12}$ |
| Fe | $(3.7 \pm 0.6) \cdot 10^{-11}$ | Ta | $< 2.0 \cdot 10^{-15}$ |
| Ga | $< 8.1 \cdot 10^{-14}$ | Th | $< 1.8 \cdot 10^{-16}$ |
| Hg | $(4.7 \pm 0.7) \cdot 10^{-13}$ | U | $< 1 \cdot 10^{-17}$ |
| In | $< 1.2 \cdot 10^{-13}$ | W | $< 4.8 \cdot 10^{-15}$ |
| K | $< 6.1 \cdot 10^{-12}$ | Zn | $(1.0 \pm 0.2) \cdot 10^{-12}$ |
| La | $< 4.3 \cdot 10^{-16}$ | | |

## 6 Inductively coupled plasma mass spectrometry

The detection methods for $^{238}$U, $^{232}$Th and $^{40}$K so far described in this paper, with the exception of the NAA, turned out not to be sensitive enough for three essential components of the BOREXINO detector and its CTF: the shielding water, the nylon for the scintillator containment vessel and the light quencher for the BOREXINO inner buffer. Besides, the NAA method is rather involved and sometimes even not available at all, because of a reactor shut-down. Alternative analytical methods thus had to be investigated. Table 10 lists the required upper limits for the concentrations of $^{238}$U, its daughter nuclides $^{226}$Ra and $^{222}$Rn, $^{232}$Th and $^{nat}$K in the CTF shielding water along with the concentrations present in the LNGS raw water. The radiopurity requirements for the nylon are at a level of $10^{-12}$ g/g for $^{238}$U and $^{232}$Th and of $10^{-9}$ g/g for K. Dimethylphthalate (DMP), the light quencher for the inner buffer, requires a detection sensitivity of $10^{-13}$ g/g for $^{238}$U and $^{232}$Th and $10^{-10}$ g/g for $^{nat}$K.

In this section we discuss measurements performed by Inductive Coupled Plasma Mass Spectrometry (ICP-MS). Multielement inorganic mass spectrometry is historically the chosen method in the laboratory for ultratrace determination due to a series of technical reasons: high degree of elemental specificity, very low detection limits together with a large dynamic range. Some limitation on its applications came essentially from its high instrumental cost combined with the high level of skill needed for operation and data interpretation. These limitations have been partially removed with the introduction of the ICP-MS in commercial form during the 80's. The detection limits obtainable with a commercial quadrupole ICP-MS are about $10^{-12}$ g/g. This is nearly sufficient for the nylon measurements and for K determinations. However, for the measurements of U and Th in the CTF shielding water the sensitivity still has to be improved.

Table 10: Summary of water radiopurity requirements.

| Element or Isotope | Raw LNGS water (Bq/kg) | CTF Design (Bq/kg) | CTF Design (g/g) |
|---|---|---|---|
| $^{238}$U | $10^{-3}$ | $10^{-6}$ | $10^{-13}$ |
| $^{226}$Ra | $3 \cdot 10^{-1}$ | $10^{-6}$ | - |
| $^{222}$Rn | 10 | $10^{-6}$ | - |
| $^{232}$Th | $10^{-3}$ | $10^{-6}$ | $2 \cdot 10^{-13}$ |
| $^{nat}$K | $10^{-3}$ | $5 \cdot 10^{-6}$ | $2 \cdot 10^{-10}$ |

## 6.1 U and Th measurements in nylon film

The best sensitivity on the U and Th impurities in nylon has been reached by using ICP-MS, combined with a careful sample preparation. This sample preparation includes a thoroughly careful cleaning with detergent and an ultrasonic bath followed by a chemical decomposition (digestion) through the addition of high purity sulfuric, nitric and perchloric acids. The digestion process turned out being fundamental for the preconcentration of nylon film and pellets. The measurements have been performed by Tama Inc., a Japanese company specialized in the production of ultra-pure chemicals (Mass Spectrometer Agilent 7500, Seiko Instruments SPQ 9000).

Table 11 reports the best values obtained so far on the candidate materials for the BOREXINO inner vessel. Durethan C38F (tradename, Bayer) is an amorphous nylon copolymer material and was used in the CTF inner vessel. The Sniamid ADS40T (tradename, Caffaro) is a nylon-6 copolymer currently under investigation. Capron B73ZP (tradename, Honeywell) is a pure nylon-6 material with no additives. Selar PA3426 (tradename, Dupont) is an engineered resin that can be mixed with nylon-6 to obtain an optically clear film.

Table 11: U and Th measured in candidate materials for the BOREXINO inner vessel.

| Material | Sample form | $^{238}$U ($10^{-12}$ g/g) | $^{232}$Th ($10^{-12}$ g/g) |
|---|---|---|---|
| Durethan C38F | Pellets | $1.2 \pm 0.4$ | $2.6 \pm 0.3$ |
| Durethan C38F | Film (Mobay) | $1.7 \pm 0.2$ | $3.9 \pm 0.5$ |
| Sniamid ADS40T | Pellets | $1.1 \pm 0.1$ | $1.6 \pm 0.1$ |
| Sniamid ADS40T | Film (Leistriz) | $2.8 \pm 0.1$ | $3.8 \pm 0.2$ |
| Capron B73ZP | Pellets | $0.46 \pm 0.04$ | $1.1 \pm 0.1$ |
| Selar PA3426 | Pellets | $0.22 \pm 0.02$ | $0.65 \pm 0.17$ |
| 90% Capron, 10% Selar | Film (Leistriz) | $1.6 \pm 0.2$ | $2.9 \pm 0.1$ |

It turned out that the cleanliness of the nylon film depends on the conditions of extrusion: a minimal contamination with dust can increase the contamination. The films measured so far have been extruded in two plants: Miles Mobay (for the inner vessel) and American Leistriz (for test purposes). In any case, the extrusion process increases the contamination of the final film (at the order of $10^{-12}$ g/g).

Another company, ACSLabs, Houston, measured a variety of cast and extruded nylon samples with similar techniques (cleaning, digestion, ICP-MS) with a sensitivity limit of $5 \cdot 10^{-11}$ g/g. Similar results have been obtained with all the rope fibers tested for the hold-down system.

## 6.2 U, Th and K measurements in water and dimethylphthalate

To reach the required levels for the shielding water (see Table 10), a HR ICP-MS (High Resolution Inductively Coupled Plasma Mass Spectrometer, VG Plasmatronic) was used in addition to all the precautions necessary for the chemical treatment of the samples. The instrument is located at the Laboratory of the Environment Institute at the Joint Research Center of the European Union (Ispra, Italy). It is assembled with a standard ICP torch-box, a sampling interface, a preliminary quadrupole focussing slit, an electrostatic analyzer (ESA), a magnetic sector and a detector system. The analyzer is practically a double focussing system based on a 70° electrostatic sector (which performs the energy focussing) and a 35° magnetic sector (which selects ions according to their mass to charge ratios). Finally, the ions are collected on a Continuous Dynode Electron Multiplier (CDEM), characteristic of a very high gain which may be further increased by coupling an Ultrasonic Nebulizer (U-5000 AT, Cetac Technologies, USA). This combination improves the limits of detection by increasing the transport efficiency and by stripping partially the solvent from the aerosol, thereby reducing the solvent load into the plasma torch. The final resulting instrumentation allows to reach a very high sensitivity ($< 10^{-14}$ g/g). However, when the starting samples are suitably preconcentrated then HR ICP-MS is potentially capable of performing measurements at the level of $10^{-15}$ to $10^{-16}$ g/g.

Samples and standards were prepared and acidified in a Class 10 clean laboratory. During the ICP-MS measurements, blank and sample containers were kept inside a special clean plastic box, shielded to avoid contaminations. Ultrapure concentrated NIST (National Institute of Standards and Technology) $HNO_3$ (70%) was used for acidification of samples and standard solutions (1/100). Usually, some reagents were redistilled three times (Berghof system, Germany) to ensure a super analytical reagent grade. LDPE (Low Density Poly Ethylene) and FEP (Fluorinated Ethylene Propylene) containers used for storage of both samples and standards were cleaned in the Class 10 clean laboratory during four weeks according to a well documented and certified procedure and checked before their use with the same measurement technique.

A very good reproducibility in the U and Th sensitivity was generally obtained even when standard solutions were prepared and analyzed at different days. Typically, using the calibration responses at ppt levels (1, 5 and 10 ppt) for plotting a linear regression, an $r^2 \cong 0.999$ value is usually obtained while the instrumental linearity is checked adding the isotope $^{112}In$ and recording the response during two hours of measurement. A variation lower then 10% is the normal result.

Different blanks (1% $HNO_3$) were analyzed in order to evaluate the U and Th concentration as well as to estimate the real limit of detection. For both the elements the Residual Standard Deviation (RSD) of replicate measurements performed on a single blank sample was comparable with the pooled RSD computed for parallel blank samples.

To assure the best quality as possible to the measurements, the so-called "standard addition" was the analytical procedure used for these applications; in others micro-aliquots of U and Th tracers at four different concentrations were added to the samples, determining through calculations the initial amount of both the nuclides present in the samples. When concentrations of U and Th isotopes were in the range of $10^{-15}$ g/g, different measurements on the same sample usually resulted in a precision between 10 and 20%. U calibration curves (optimized with the code Minuit) typically gave a sensitivity between 800 and 1100 counts for a concentration of $10^{-12}$ g/g (depending on whether or not the Ultrasonic Nebulizer was in use). Using an integration time of 8 s, the Limit Of Detection (3σ) was $10^{-15}$ g/g.

Table 12 shows values which are typical for the ICP-MS measurements of the CTF water made over several years. They refer to the production line (Gran Sasso water processed once by the purification system), to the main tank (water residing in the CTF storage tank), and to the circulation loop (water that was in the main tank and has just passed the recirculation/purification system). The results are consistent with the requirements. Despite of the continuous treatment of the water by means of a recirculation loop [26], the level of $^{222}$Rn (20-30 mBq/m$^3$) in the CTF water was higher than the equivalent $^{226}$Ra contamination (< 2 mBq/m$^3$) due to the $^{222}$Rn emission from the CTF materials as explained above. Then it was very important to have almost continuous monitoring, which has been carried out via ICP-MS measurements. Some measurements were also performed on the water taken directly inside the CTF tank, to assess the level of water recontamination in the CTF.

Table 12: Typical values (in g/g) for $^{238}$U, $^{232}$Th and $^{nat}$K in water measured with ICP-MS at Ispra (see text).

| Element | Production Line | Circulation Loop | Main Tank |
|---|---|---|---|
| $^{238}$U | $10^{-14}$ | $10^{-14}$ | $3 \cdot 10^{-14}$ |
| $^{232}$Th | $10^{-14}$ | $10^{-14}$ | $3 \cdot 10^{-14}$ |
| $^{nat}$K | $5 \cdot 10^{-15}$ | $5 \cdot 10^{-15}$ | $10^{-14}$ |

The same technique has been used to assay the quality of DMP, the organic liquid used as a quencher in the Borexino inner buffer, yielding $5 \times 10^{-14}$ g/g for $^{238}$U and $^{232}$Th and $2 \times 10^{-10}$ g/g for $^{nat}$K. The potassium result has been independently confirmed by ChemTrace, a U.S. company which performed for us measurements with the cool plasma ICP-MS technique [27].

## 7   Graphite furnace atomic absorption spectroscopy

For the measurement of K in nylon an additional detection method has been applied. This is because the measurement of K with NAA (section 5) is limited by the presence of sodium in the sample, while the measurement with ICP-MS (section 6) is impaired by the use of argon. Tests by means of NAA (Technical University Munich, Missouri University Research Reactor) and by means of ICP-MS-Axial (ACSLabs) have been carried out; however, the best sensitivity has been reached by Tama Inc. company through Graphite Furnace Atomic Absorption Spectroscopy (GFAAS) using a Perkin Elmer PE5100 spectrometer.

The operating principle of GFAAS is based on the correlation between the intensity of absorbed light and the concentration of free atoms of any specific element. Samples are deposited in small graphite tubes and brought to high temperatures in order to vaporize and atomize them. Light at wavelengths specific to the element of interest passes through the tube, where it will be absorbed by free atoms produced during the heating.

The results of the GFAAS measurements are summarized in Table 13 along with measurements performed by NAA and ICP-MS. The results from the same sample obtained with different methods and in different laboratories are sometimes not consistent; this is probably due to the difficulty to maintain the cleanliness of the measured sample during the processing.

Table 13: Summary of K measurements in nylon samples using different methods.

| Material | K ($10^{-9}$ g/g) | Method |
|---|---|---|
| C38F pellets | $0.6 \pm 0.2$ | GFAAS (Tama) |
| C38F film | $2.9 \pm 0.5$ | GFAAS (Tama) |
| ADS40T pellets | 1.6 | NAA (Munich) |
| ADS40T pellets | 14 | GFAAS (Tama) |
| ADS40T pellets | $25 \pm 9$ | NAA (Missouri) |
| Capron pellets | 14 | GFAAS (Tama) |
| Capron pellets | 25 | NAA (Munich) |
| Capron pellets | $13 \pm 6$ | NAA (Missouri) |
| Capron pellets | < 5 | ICPMS-Axial (ACS) |
| Capron film | $5 \pm 1$ | GFAAS (Tama) |

# 8 The Counting Test Facility

The study for feasibility of BOREXINO has been carried out by means of the Counting Test Facility (CTF), a simplified and scaled version of the BOREXINO detector, installed in Hall C of the Gran Sasso Underground Laboratory. An array of 100 phototubes, coupled to optical concentrators, detects the light emitted from 4.8 m$^3$ of liquid scintillator, contained in a transparent spherical nylon vessel with 1.05 m radius. A ~4.5 m shielding layer is provided in all directions by 1000 m$^3$ of high purity water (see section 6). The scintillators under study were PC/PPO and PXE/pTP/bis-MSB (as described in section 5.5).

The measured quantities in CTF and in BOREXINO are: (1) a calorimetric measurement of the energy with a typical resolution of 10%; (2) the position of the events in space derived from the phototube timing; (3) $\alpha/\beta$ particle identification to discriminate against background; and (4) the time correlation between events to disentangle the correlated decays of background events. For a wider description see [2].

All the construction materials have been selected to keep their radioactive content below an acceptable limit (see Table 14); the selection was carried out by means of high sensitivity germanium $\gamma$-ray spectrometers (section 3) and NAA (section 5). The choice of materials with low contamination assures not only a low direct emission of $\gamma$ radiation, but also a low $^{222}$Rn emanation.

Table 14: Summary of the Counting Test Facility (CTF) radiopurity results.

| Material | Radio purity | Achieved in CTF |
|---|---|---|
| Stainless steel | Th, U equiv. | ~ $10^{-9}$ g/g |
| Shielding water | " " | ~ $10^{-14}$ g/g |
| Photo multipliers | " " | ~ $10^{-8}$ g/ |
| Nylon film | " " | ~ $10^{-12}$ g/g |
| Buffer liquid | " " | ~ $10^{-15}$ g/g |
| Scintillator | " " | ~ $4 \cdot 10^{-16}$ g/g |
| " | $^{14}$C/$^{12}$C | ~ $2 \cdot 10^{-18}$ |

Due to the high $^{222}$Rn level in Hall C of the Gran Sasso Laboratory (40-80 Bq/m$^3$, depending on the ventilation) the auxiliary plants have to be not only air-tight, but also fully impermeable to Rn; this demand has an impact on the choice of materials, on their thickness as well as on the gaskets used. The surfaces have to be treated at "electronic grade", i.e. smoothed (0.6-0.8 μm), pickled and passivated or electro-polished, to provide a clean surface which in turn allows an easy cleaning for dust and particles.

The spherical nylon vessel has been built using an amorphous nylon copolymer resin, Durethan C38F (described above). The U and Th content of the pellets was at a level of a few times $10^{-12}$ g/g (see Table 11). The full installation has been done in a controlled area, ranging from class 10 to class 10000. All the methods and techniques, tested and applied to the CTF, are presently employed in the construction of the full scale BOREXINO detector. The total count rate of the CTF is 0.5 counts keV$^{-1}$ kg$^{-1}$ y$^{-1}$ in the range of 250-2500 keV. Defining only part of the inner vessel as fiducial volume can reduce the background even further.

The CTF has been used until now to measure the radiocontamination of scintillators based on PC and PXE. These studies included in particular the radioactive nuclides from the U and Th families, as well as $^{14}$C. The activity of the latter has been derived from an analysis of the energy spectrum below 250 keV. In this range the count rate is dominated by the $^{14}$C activity. The measured value expressed as a ratio $^{14}$C/$^{12}$C yielded $(1.94 \pm 0.09) \cdot 10^{-18}$ for PC [3] and $(11.74 \pm 0.03) \cdot 10^{-18}$ for PXE [28]. The $^{238}$U and $^{232}$Th equivalent rates have been measured via the delayed coincidences $^{214}$Bi-$^{214}$Po with a lifetime of 236 µs and $^{212}$Bi-$^{212}$Po with a lifetime of 0.433 µs, respectively [4]. In the case of the $^{238}$U family only an upper limit could be determined, since $^{214}$Bi and $^{214}$Po are daughter nuclides of $^{222}$Rn, which permeates from the water through the wet nylon into the scintillator. Therefore, the $^{222}$Rn activity expressed as $^{238}$U equivalent is $(3.5 \pm 1.3) \cdot 10^{-16}$ g/g for PC [4] and $(7.3 \pm 2.4) \cdot 10^{-16}$ g/g for PXE [28]. The difference in the $^{222}$Rn level (a factor of 2) between PC and PXE can be explained by an altered detector performance. A direct $^{238}$U assay via NAA (see section 5.5 and Tables 8, 9) yielded distinct upper limits, in particular for PXE ($<10^{-17}$ g/g). The $^{232}$Th equivalent rate measured with the CTF was $(4.4^{+1.5}_{-1.0}) \cdot 10^{-16}$ g/g for PC [4] and $(4.0 \pm 4.4) \cdot 10^{-16}$ g/g for PXE [29]. Again, the directly measured limit for PXE obtained with NAA is even better ($<1.8 \cdot 10^{-16}$ g/g, see Table 9). We wish to emphasize that these numbers are records in low radioactivity measurements, they are well below the limits obtainable even with the most sensitive mass spectrometers (see section 6).

The sensitivity of the CTF could possibly be further improved by installing a nylon balloon as barrier against the $^{222}$Rn emanated by the various components of the CTF (the PMTs, the internal walls of the water containment tank, etc). A gain by an order of magnitude in the total sensitivity and probably by some factor in the sensitivity within the inner vessel after the spatial reconstruction is expected from this improvement.

Very low background detectors (VLBD) could be successfully run by installing them in the center of the CTF, which in this way serves as a very good shielding system. If the CTF is used only as a passive shield, it would be entirely filled with water and the PMTs would be dismounted (though their contribution to the background is almost negligible). A VLBD with a volume included in a 40 cm sphere placed at the center of the detector would be in a low background environment at a level of 14 counts/day in the range of 250-500 keV and 60 counts/day in the range 0-3500 keV, respectively. These backgrounds could be expected to be pushed down to about 2 counts/day and 7 counts/day, respectively, when the nylon barrier against $^{222}$Rn will be installed.

Finally, the CTF could be used also to measure the radioactivity of different objects, which can be installed at its center. In this case the active scintillator volume and the PMTs are needed. The external background (with the $^{222}$Rn barrier installed) is dominated by $^{222}$Rn and its contribution to the count rate is the same as quoted above. However, the total sensitivity depends on the kind of events which have to be detected and whether or not they can be tagged.

This capability of the CTF is of key interest for future potential applications and could certainly motivate the construction of a second one for low background measurement purposes.

The radioactive contamination of material samples can also be measured simply by introducing them into the shielding water of the CTF, very close to the inner vessel containing the scintillator, without modifications of the detector. In this case the sensitivity is limited by the total background quoted above.

## 9   Conclusions

The needs of the BOREXINO solar neutrino experiment required the knowledge of ultra-low levels of radioactivity and in many cases, the development of new techniques in chemical preparations and radioisotope counting. In this paper, we presented the results of specific measurements on a number of materials of importance not only to our experiment but also to other experiments in Particle Astrophysics. In addition we described our own new techniques and instrumental developments. Here we summarize the main points.

We operate a number of underground Ge detectors, whose backgrounds are among the lowest achieved so far. Such detectors allowed the measurement of $10^{-9}$-$10^{-11}$ g/g of $^{238}$U, $^{232}$Th, $^{226}$Ra and few $10^{-9}$ g/g of K as lowest radioactive contents in various construction materials for BOREXINO.

State of the art sensitivity in $^{222}$Rn collection and measurement has been achieved through our developments in miniaturized proportional detectors and $^{222}$Rn extraction and trapping. These techniques have evolved in instruments used to measure $^{222}$Rn in water and $^{222}$Rn emanation of materials and in large volumes (such as storage tanks). In particular, the $^{222}$Rn emanation measurements can reach a sensitivity of 10 μBq/m$^2$ for large surface samples. The $^{226}$Ra concentration in water can be determined with a sensitivity of ~1 mBq/m$^3$, for $^{222}$Rn it is 0.1 mBq/m$^3$. $^{222}$Rn in gases, and particularly in N$_2$, can be measured with a sensitivity below 1 μBq/m$^3$.

Developments in low-level scintillator counting, a key feature of BOREXINO, have led us to a number of new techniques. These include the development of the Counting Test Facility in order to achieve the ultra-low measurements of $^{14}$C and $^{232}$Th at the levels required for BOREXINO. In particular, the CTF could measure the $^{14}$C/$^{12}$C ratio down to $10^{-18}$ g/g. Limits on $^{226}$Ra (based on $^{222}$Rn) and $^{232}$Th were established to the $10^{-16}$ g/g level. The $^{232}$Th and $^{238}$U concentrations in aromatic liquid compounds were measured studying the delayed coincidence patterns of the pairs $^{214}$Bi-$^{212}$Po and $^{212}$Bi-$^{212}$Po in the respective radioactive chains.

The techniques of low-level counting have been combined with Neutron Activation Analysis (NAA). The best results are obtained using the β-γ coincidences of carefully processed samples following reactor irradiation. NAA studies the heads of the natural decay families in particular. These levels (down to $10^{-17}$ g/g of U and Th) are very critical for the central scintillator in BOREXINO and cannot be measured directly by the CTF which uses the Bi-Po pairs method. NAA has been also used to measure the K in the inner vessel materials to the $10^{-9}$ g/g range. The NAA techniques included the ability to irradiate liquid scintillator samples in the reactor pool itself.

Techniques in sample handling and chemistry have been developed to process samples using the commercially available mass spectrometer and atomic absorption systems. The isotopes of $^{238}$U, $^{232}$Th, $^{40}$K can be measured in water by means of the ICP-MS down to few $10^{-15}$ g/g. ICP-MS has been used also to determine the U and Th contents of the candidate nylon compounds for the scintillator containment vessel, reaching a sensitivity in the $10^{-12}$ g/g range. The K contamination in nylon, in the fluor and in the quencher has been measured with a variety of methods, as GFAAS, ICP-MS-axial and cold plasma ICP-MS with sensitivities in the $10^{-9}$ g/g range.

The techniques and results reported in this paper fulfill the needs of BOREXINO. We believe that the measured activity levels of the materials listed in the tables will be useful for the design of future experiments in Particle Astrophysics.

**Acknowledgements**.

We sincerely thank the funding agencies: Istituto Nazionale di Fisica Nucleare [INFN] (Italy); National Science Foundation [NSF] (USA); Bundesministerium für Bildung, Wissenschaft, Forschung und Technologie [BMBF], Deutsche Forschungsgemeinschaft [DFG] (Germany). We acknowledge the very generous support given to our work by the Laboratori Nazionali del Gran Sasso [LNGS] (Italy).

**References**

[1]  G. Alimonti et al., Borexino collaboration, "Science and Technology of Borexino", accepted for publication in Astroparticle Phys. (2001) (hep-ex/0012030); F. Calaprice et al., Borexino Collaboration, "Proposal for a real time detector for low energy solar neutrinos", Ed. by the Princeton University (1992); C. Arpesella et al., Borexino Collaboration, "Borexino at Gran Sasso: Proposal for a real time detector for low energy solar neutrinos", G. Bellini, M. Campanella, D. Giugni, and R. Raghavan (Eds.) INFN Milano (1991).

[2]  G. Alimonti et al, Borexino Collaboration, "A large scale low background liquid scintillator detector: The counting test facility at Gran Sasso", Nucl. Instr. Meth. A406 (1998) 411.

[3]  G. Alimonti et al., Borexino Collaboration, "Measurement of the $^{14}$C abundance in a low background liquid scintillator", Phys. Lett. B422 (1998) 349.

[4]  G. Alimonti et al., Borexino Collaboration, "Ultra-low background measurements in a large volume underground detector", Astroparticle Phys. 8 (1998) 141.

[5]  E. Bellotti et al., "New measurement of rock contaminations and neutron activity in the Gran Sasso tunnel", INFN/TC-85/19; P. Belli et al., Il Nuovo Cimento 101A (1989) 959.

[6]  C. Arpesella, "A low background counting test facility at Laboratori Nazionali del Gran Sasso", Appl. Rad. Isot. 47 (1996) 991.

[7]  H. Neder, G. Heusser, M. Laubenstein, Appl. Rad. Isot. 53 (2000) 181.

[8]  M. Hult et al., Appl. Rad. Isot. 53 (2000) 225.

[9]  G. Heusser, Nucl. Inst. Meth. B58 (1991) 79.

[10]  M. Wójcik et al., Nucl. Inst. Meth. A449 (2000) 158.

[11]  R. Wink et al., "The miniaturized proportional counter HD-2(Fe)/(Si) for the Gallex solar neutrino experiment", Nucl. Instr. Meth. A329 (1993) 541.

[12]  M. Laubenstein, "Messungen von $^{222}$Rn und $^{226}$Ra im Rahmen der Counting Test Facility des Sonnenneutrinoexperiments Borexino", PhD thesis, Universität Heidelberg (1996).

[13]  W. Rau, "Low-Level-Radonmessungen für das Sonnenneutrinoexperiment Borexino", PhD thesis, Universität Heidelberg (1999).

[14]  W. Rau, G. Heusser, "$^{222}$Rn emanation measurements at extremely low activity", Appl. Rad. Isot. 53 (2000) 371.

[15]  H. Simgen, "Messung von $^{222}$Rn und $^{226}$Ra in Wasser im Rahmen des Sonnenneutrino-Experiments Borexino", Diploma thesis, Universität Heidelberg (2000).

[16] G. Heusser et al., "$^{222}$Rn detection at the microBq/m$^3$ range in nitrogen gas and new $^{222}$Rn purification technique for liquid nitrogen", Appl. Rad. Isot. 52 (2000) 691.

[17] B. Freudiger, "Bestimmung des Radon Gehaltes in flüssigem Stickstoff", Diploma thesis, Universität Heidelberg (1998).

[18] J. Kiko, Nucl. Instr. Meth. A460 (2001) 272.

[19] M. Schuster, S. Ringmann, R. Gärtner, X. Lin, J. Dahmen., Fres. J. Anal. Chem, **357**(3) (1997) 258.

[20] R. Raghavan, AT&T, Technical Report, Oct. 1992.

[21] T. Goldbrunner et al., "Neutron activation Analysis of Detector Components for the Solar Neutrino Experiment BOREXINO", J. Radioanal. Nuc. Chem. **216**(2) (1997) 293.

[22] T. Goldbrunner et al., "New Records on Low Activity Measurements with Neutron Activation Analysis", Nucl. Phys. B, (Proc. Suppl.) 61B (1998) 176, Proc. of "International Conference on Advanced Technology and Particle Physics", Como, 7 - 11 October 1996.

[23] R. v. Hentig et al., Fres. J. Anal. Chem. 360 (1989) 664.

[24] L.A. Currie, Anal. Chem., **40**(3) (1968) 586.

[25] J. Kim et. al., Nucl. Instr. Meth., 177 (1980) 557.

[26] M. Balata et al., "The water purification system for the low background counting test facility of the Borexino experiment at Gran Sasso", Nucl. Instr. Meth. A 370 (1996) 605; M.G. Giammarchi et al., Ultrapure Water Journal 13,8 (1996) 59.

[27] D. Wollenweber et al., "Determination of Li, Na, Mg, K, Ca and Fe with ICP-MS using cold plasma conditions", Fres. J. Anal. Chem. 364 (1999) 433; S. Georgitis et. al.,"'Cool' Plasma technique for the ultra-trace level determination of $^{56}$Fe, $^{40}$Ca, $^{39}$K, $^{7}$Li, $^{24}$Mg and $^{23}$Na for Semi-Conductor applications", Varian Inc. online resource (www.varianinc.com/osi/icpms/atwork/icpms7.htm).

[28] E. Resconi, "Measurements with the Upgraded Counting Test Facility (CTF-2) of the Solar Neutrino Experiment Borexino", PhD thesis, Universitá degli Studi di Genova (2001).

[29] M. Göger-Neff, "Developments for the data analysis and for a liquid scintillator for the solar neutrino experiment Borexino", PhD thesis, Technische Universität München (2001).